%% file: ms.tex
\documentclass{pasj01}

\usepackage{times}
\usepackage{color}
\usepackage{comment}
\usepackage{lscape}
\usepackage{natbib,aas_macros}
\newcommand{\Lya}{Ly$\alpha$}
\newcommand{\OII}{[O\,\textsc{ii}]}
\newcommand{\Ha}{H$\alpha$}
\newcommand{\Hb}{H$\beta$}
\newcommand{\Hc}{H$\gamma$}
\newcommand{\Hd}{H$\delta$}

\newcommand{\NII}{[N\,\textsc{ii}]}
\newcommand{\NIIIsemi}{N\,\textsc{iii}]}
\newcommand{\OIII}{[O\,\textsc{iii}]}
\newcommand{\OIIIsemi}{O\,\textsc{iii}]}
\newcommand{\SII}{[S\,\textsc{ii}]}
\newcommand{\HeII}{He\,\textsc{ii}}
\newcommand{\AlII}{Al\,\textsc{ii}}
\newcommand{\NeIII}{[Ne\,\textsc{iii}]}

\newcommand{\HII}{H\,\textsc{ii}}

\newcommand{\COSMOS}{ID-1} 
\newcommand{\SMACS}{ID-2} 
\newcommand{\BXI}{ID-3} 
\newcommand{\BXII}{ID-4} 
\newcommand{\BXIII}{ID-5} 
\newcommand{\Lynx}{ID-6} 
\newcommand{\AbellYK}{ID-7} 
\newcommand{\SGAS}{ID-8} 
\newcommand{\CSWA}{ID-9} 
\newcommand{\MACS}{ID-10} 
\newcommand{\AbellS}{ID-11} 
\newcommand{\KBSS}{S16-stack} 

\newcommand{\COSMOSno}{COSMOS\,12805} 
\newcommand{\SMACSno}{SMACS\,0304} 
\newcommand{\BXIno}{BX74} 
\newcommand{\BXIIno}{BX418} 
\newcommand{\BXIIIno}{BX660} 
\newcommand{\Lynxno}{the Lynx arc} 
\newcommand{\AbellYKno}{Abell\,22.3} 
\newcommand{\SGASno}{SGAS\,1050} 
\newcommand{\CSWAno}{CSWA\,20} 
\newcommand{\MACSno}{MACS\,0451} 
\newcommand{\AbellSno}{Abell\,860\_359} 

\newcommand{\q}{$q_{\rm ion}$}

\Received{24-Nov-2016}
\Accepted{24-Feb-2017 (in press)}

\begin{document}

\SetRunningHead{T. Kojima et al.}{Evolution of N/O and $q_{\rm ion}$ at $z\sim 0$--$2$}

\title{
Evolution of N/O Abundance Ratios and Ionization Parameters from $\lowercase{z}\sim 0$ to $2$ Investigated by the Direct Temperature Method
\altaffilmark{\dag}
\altaffilmark{\ddag}
}

\author{
Takashi Kojima \altaffilmark{1,2},
Masami Ouchi \altaffilmark{1,3},
Kimihiko Nakajima \altaffilmark{4},\\
Takatoshi Shibuya \altaffilmark{1},
Yuichi Harikane \altaffilmark{1,2}, and
Yoshiaki Ono \altaffilmark{1}
}

\email{tkojima@icrr.u-tokyo.ac.jp}
\altaffiltext{1}{%
Institute for Cosmic Ray Research, The University of Tokyo,
5-1-5 Kashiwanoha, Kashiwa, Chiba 277-8582, Japan
}
\altaffiltext{2}{%
Department of Physics, Graduate School of Science, The University of Tokyo, 7-3-1 Hongo, Bunkyo, Tokyo 113-0033, Japan
}
\altaffiltext{3}{%
Kavli Institute for the Physics and Mathematics of the Universe (WPI),
The University of Tokyo, 5-1-5 Kashiwanoha, Kashiwa, 
Chiba 277-8583, Japan
}
\altaffiltext{4}{%
European Southern Observatory, Karl-Schwarzschild-Strasse 2, 
D-85748 Garching, b. M\"unchen, Germany
}

\altaffiltext{\dag}{%
Some of the data presented herein were obtained at the W.M. Keck Observatory, 
which is operated as a scientific partnership among the California Institute 
of Technology, the University of California and the National Aeronautics and 
Space Administration. The Observatory was made possible by the generous 
financial support of the W.M. Keck Foundation.
}
\altaffiltext{\ddag}{%
Based in part on data collected at Subaru Telescope,
which is operated by the National Astronomical Observatory of Japan.
}


\maketitle

\begin{abstract}
We present N/O abundance ratios, ionization parameters {\q}, and oxygen abundance O/H for 
a total of 41 galaxies (11 individual galaxies and a 30-galaxy stack) including {\Lya} emitters and Lyman break galaxies at $z\sim 2$,
and investigate galaxy evolution from $z\sim 0$ to $2$ in conjunction with
208,529 local SDSS galaxies
and 9 green pea galaxies (GPs).
In contrast with most of the recent studies,
we obtain the N/O ratio, {\q}, and O/H measurements 
by direct $T_{\rm e}$ methods 
with {\OIII} $\lambda$4363 and {\OIIIsemi} $\lambda$1665 lines. 
Based on these reliable measurements,
we find that there exist $z\sim 2$ galaxies with an excess of N/O falling beyond
the local average of N/O--O/H relation, 
while the majority of the $z\sim 2$ galaxies have the N/O ratios nearly comparable with $z\sim 0$ galaxies in the N/O--stellar mass relation.
Our galaxies place the upper limit of N/O ratio $\log ({\rm N/O})\le -1.26$ on average, suggesting that 
the N/O ratio evolves, if at all, by $<0.17$ dex. 
Exploiting our reliable measurements free from the N/O--{\q}--O/H degeneracies,
we identify, for the first time, that $z\sim 2$ galaxies with offsets in the BPT diagram
indicate either 1) only an N/O excess, 2) only a {\q} excess, or 
3) both N/O and {\q} excesses. 
We argue that 
the BPT offsets at $z\sim 2$ are not made by one of 1)--3) galaxy populations alone,
but the composite of 1)--3) populations.
We confirm that these 1)--3) populations also exist
at $z\sim 0$, such as GPs and SDSS low-mass and high-SFR galaxies (LMHSs).
\end{abstract}

\section{INTRODUCTION} \label{sec:intro}
The {\OIII} $\lambda$5007/{\Hb} vs. {\NII} $\lambda$6584/{\Ha} diagram, often referred to as the Baldwin--Phillips--Terlevich \citep[BPT][]{Baldwin1981} diagram, is used to distinguish star-forming galaxies (SFGs) from active galactic nuclei (AGNs). 
Local SFGs make a tight relation in the BPT diagram, which is caused by the anti-correlation between the oxygen abundance, 12+log(O/H), and the ionization parameter, {\q} \citep{Dopita1986, Dopita2000, Dopita2006a}.
Recent studies have revealed that $z\sim2$ galaxies have {\NII} $\lambda$6584/{\Ha} (N2) values higher than those of the local SFGs for a given {\OIII} $\lambda$5007/{\Hb} (O3) on average \citep{Shapley2005b, Erb2006a, Brinchmann2008a, Liu2008, Kewley2013b, Steidel2014, Masters2014, Yabe2014, Hayashi2015, Shapley2015, Sanders2016a, Kashino2016}. The $z=2$ galaxies observed in the Keck Baryonic Structure Survey \citep[KBSS;][]{Steidel2014} show the average of $\sim 0.4$ dex excesses in {\NII} $\lambda$6584/{\Ha}. These differences between the local and high-$z$ SFGs are called ``BPT offsets''.

The BPT offset indicates the evolution of the physical conditions of the interstellar medium (ISM) in galaxies from $z\sim0$ to $z\sim2$.
To explain the BPT offset,
there are four possible scenarios of galaxy evolution towards $z\sim 2$: high electron densities $n_e$, 
high nitrogen-to-oxygen abundance ratios N/O,
high ionization parameters, and hard far-ultra violet (FUV) radiation fields.
Photoionization models suggest that 
the BPT offsets appear, if, at least, one of the four parameters
is changed from values of typical local galaxies
\citep[e.g.][]{Kewley2013a, Steidel2014, Sanders2016a}.

For the explanation of the $n_e$ redshift evolution, the typical $n_e$ values of $z=2$--$3$ SFGs are estimated with {\OII} $\lambda \lambda$3727/3729 and/or {\SII} $\lambda \lambda$6716/6731 ratios to be $n_e\sim250$ cm$^{-3}$ \citep{Steidel2014, Shimakawa2015, Sanders2016a}. This $n_e$ value at $z=2$--$3$ is about an order of magnitude higher than the typical value of local galaxies, $n_e\sim25$ cm$^{-3}$, which is derived by \citet{Sanders2016a} from the sample of the Sloan Digital Sky Survey \citep[SDSS;][]{York2000}.
Although there is a large difference of the $n_e$ values 
between $z=2$--$3$ and local SFGs, 
photoionization models indicate that this $n_e$ difference can increase {\OIII} $\lambda$5007/{\Hb} and {\NII} $\lambda$6584/{\Ha} values only by $\sim$ 0.01, 0.05, and 0.1 dex for SFGs with the oxygen abundances of 12+log(O/H)=8.0, 8.3, and 8.5, respectively \citep{Sanders2016a}. 
These small changes 
cannot
explain the BPT offset, 
suggesting that the $n_e$ evolution is not the major source of the BPT offset.

For the explanation of the N/O redshift evolution, \citet{Queyrel2009}, \citet{Masters2014}, and \citet{Yabe2015} have estimated N/O ratios at $z\sim2$, and have claimed the increase of N/O at fixed O/H. 
The excessive N/O ratios at fixed O/H may be caused by the decrease of O/H due to the pristine gas inflow into galaxies with rich nitrogen gas produced by the secondary nucleosynthesis, 
while Wolf-Rayet (WR) stars can produce the nitrogen in a time scale shorter than the oxygen production time scale in supernovae \citep{Andrews2013}.
Galaxies with an excessive N/O ratio tend to indicate the WR-star feature in their spectrum \citep[e.g.,][]{Pustilnik2004, Perez-Montero2011}.
Here, \citet{Queyrel2009} and \citet{Yabe2015} have derived the N/O values based on the $z\sim0$ empirical relation with the ratios of the strong emission lines of {\NII} $\lambda$6584/{\SII} $\lambda\lambda$6716,6731 (N2S2). 
This N2S2 index is calibrated
by \citet{Perez-Montero2009} and \citet{Amorin2010}. 
\citet{Masters2014} derive the N/O ratio, 
assuming that the electron temperature ($T_e$) is 10,000 K.
However, with local SDSS galaxies, the strong line method gives N/O values systematically higher than the reliable direct temperature ($T_e$) method (see Section \ref{sec:result} for more details).
Moreover, there remain possible systematics in applying this $z\sim 0$ empirical N2S2 relation to $z\sim2$ SFGs
whose ISM properties should be different from $z\sim 0$ galaxies.

For the explanations of the {\q}, \citet{Nakajima2014} have reported that the {\q} values of SFGs are systematically higher at $z=2$--$3$ than at $z\sim0$
by an order of magnitude. 
The values of 12+log(O/H) and {\q} are estimated with the combination of the strong line
indicators of R$_{23}$ and O$_{32}$ defined as
\begin{eqnarray}
\textrm{R$_{23}$} &\equiv&\log \left(\frac{{\textrm{\OII}\,\lambda3727}+{\textrm{\OIII}\,\lambda\lambda4959,5007}}{\textrm{H}\beta}\right) \\
\textrm{O$_{32}$} &\equiv&\log \left(\frac{\textrm{\OIII}\,\lambda\lambda4959,5007}{\textrm{\OII}\,\lambda3727}\right),
\end{eqnarray}
which were introduced by \citet{Kobulnicky2004}.
For the standard {\HII}-region model of a Str\"omgren sphere, the dependence of {\q} on $n_e$ and hydrogen photoionizing photon production rate $Q_{\rm H0}$ is given by $q_{\rm ion} \propto (Q_{\rm H0} n_e)^{1/3}$ \citep{Charlot2001,Shirazi2014b,Nakajima2014}.
The factor of 10 evolution in the electron density is not enough to be the major cause of the {\q} evolution.
Thus, the redshift evolution of {\q} would be largely contributed by the increase of the 
hydrogen photoionizing production rate 
in galaxies from $z\sim 0$ to $z\sim 2$. 
The photoionizing rate and the radiation hardness are determined by stellar population properties including stellar age, metallicity, dust extinction, and the initial mass function (IMF).
The physical origins of {\q} and the radiation hardness largely overlap, and produce similar BPT offsets.
In this sense, the independent determinations of {\q} and radiation hardness are challenging in observations, which means the {\q} and radiation hardness have a degeneracy.
We thus do not investigate the degeneracy of {\q} and radiation hardness further in this paper, but instead assume the non-evolving radiation hardness.

Because the $n_e$ evolution cannot explain the BPT offset under the assumption of the non-evolving radiation hardness, in this study, we investigate the remaining two possibilities, the evolution of N/O and {\q}.
Our goal is to understand the N/O and {\q} values of $z\sim 2$ galaxies based on the direct $T_e$ method.
We use multiple line ratios including {\OIII} $\lambda$4363 and {\OIIIsemi} $\lambda\lambda$1661,1666 for the reliable $T_e$ method to derive 12+log(O/H), N/O, and {\q} values that are free from many of the systematic uncertainties explained above. 
For observational data of this study, we conducted rest-frame infra-red (IR) spectroscopy targeting a $z\sim2$ {\Lya} emitter (LAE), 
{\COSMOSno}, which has a significant detection of the {\OIIIsemi} $\lambda\lambda$1661,1666 lines useful for the direct $T_e$ method estimates \citep{Shibuya2014b}.
We also use ten $z\sim2-3$ galaxies with {\OIII} $\lambda$4363 or {\OIIIsemi} $\lambda\lambda$1661,1666 measurements from the literature and the composite spectrum made of thirty $z\sim2$ galaxies by \citet{Steidel2016}.

The outline of this paper is as follows.
In Section \ref{sec:data}, we describe our data and observations of {\COSMOSno}.
Section \ref{sec:sample} presents our sample of $z\sim2$ galaxies, and discusses the reliability of the galaxies in our sample.
In Section \ref{sec:analysis}, we explain the measurements of $T_e$, O/H, N/O, and {\q} of the galaxies in our sample.
In Section \ref{sec:result}, we discuss three relations of N/O--$M_{\star}$, N/O--O/H, and {\q}--O/H for the $z\sim2$ galaxies. Here we discuss how these three relations are relevant to the BPT diagram offset.
We summarize our results in Section \ref{sec:summary}.
Throughout this paper, magnitudes are on the AB system \citep{Oke1983}. 
We adopt the following cosmological parameters, $(h, \Omega_m, \Omega_{\Lambda}) = (0.7, 0.3, 0.7)$.

\section{DATA AND OBSERVATIONS} \label{sec:data}
One of our sample SFGs, {\COSMOSno}, is an LAE at $z=2.159$ (RA$=10^{\rm h}00^{\rm m}15.^{\rm s}29$, Dec$=+02^{\circ}08^{\prime}07.^{\prime\prime}48$).
This object is selected by the excess of narrowband $NB387$ flux in the two color diagram of 
$B-NB387$ vs. $u^{*}-NB387$ \citep{Nakajima2012, Nakajima2013}.
\citet{Shibuya2014b} have carried out follow-up spectroscopic observations for {\COSMOSno} with the Low Resolution Imaging Spectrometer \citep[LRIS; ][]{Oke1998, Steidel2004} on the Keck-I telescope, 
and detected {\OIIIsemi} $\lambda\lambda$1661,1666 emission lines in the spectrum of {\COSMOSno}.

In addition, our observation with the Subaru Fiber Multi Object Spectrograph \citep[FMOS; ][]{Kimura2010} gives constraints 
on nebular line fluxes of {\OII} $\lambda$3727, {\Hb}, and {\OIII} $\lambda\lambda$4959,5007. More recently, our program with 
the Multi-Object Spectrometer For Infra-Red Exploration \citep[MOSFIRE; ][]{McLean2010, McLean2012} on the Keck-I telescope 
covers emission lines of {\Ha}, and {\NII} $\lambda$6584. In this section, we describe these observations 
and data
of the Keck/LRIS, 
Subaru/FMOS, and Keck/MOSFIRE.

\input tab1.tex

\subsection{LRIS Data} \label{subsec:lris}
The Keck/LRIS observations were carried out in 2012 March 19--21 (UT) 
with seeing sizes of $1.\!\!^{\prime\prime}0$--$1.\!\!\arcsec4$ (FWHM).
The total exposure time was 24,000 seconds for the mask targeting {\COSMOSno}.
The spectral resolution is $R\sim1000$, which corresponds to $\simeq300$ km s$^{-1}$. Details of our LRIS observations and data reduction are presented in \citet{Shibuya2014b}. 

\begin{figure}[t!]
\begin{center}
\includegraphics[scale=0.35, bb=0 0 651 873]{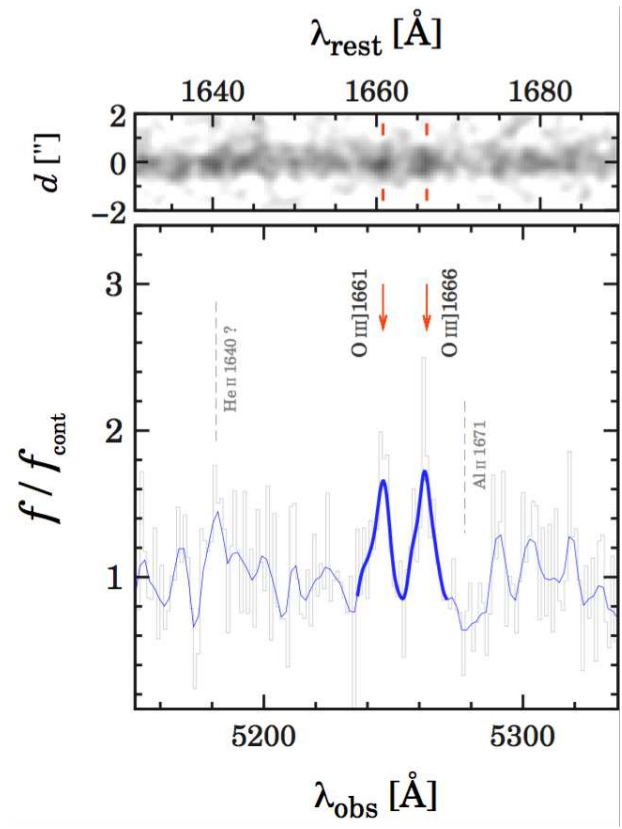}
\end{center}
\caption{{\footnotesize Rest-frame UV spectrum around the {\OIIIsemi} doublet of {\COSMOSno}. The top and bottom panels show the two- and one-dimensional (2D and 1D) spectra, respectively. In the bottom panel, the gray solid line indicates the spectrum normalized to unity in the UV continuum level. The blue solid line presents a spectrum smoothed with a Gaussian function of $\sigma=3$ pix. The red arrows and the gray vertical dashed lines denote expected wavelengths of inter-stellar emission (i.e. {\OIIIsemi} and {\HeII} $\lambda$1640) and absorption (i.e. {\AlII} $\lambda$1671) lines for the systemic redshift of $z_{\rm sys}=2.159$ (see Section \ref{subsec:fmos}). }}
 \label{fig:spec_uv}
\end{figure}

Figure \ref{fig:spec_uv} shows the rest-frame UV spectrum normalized by the UV continuum flux density of {\COSMOSno}. The {\OIIIsemi} $\lambda$1661 and {\OIIIsemi} $\lambda$1666 lines (refereed to as {\OIIIsemi} $\lambda$1665 line) are detected
at signal-to-noise (S/N) ratios of $4.0$ and $5.1$, respectively.
Based on the wavelength interval and the $S/N$ ratios of the two lines,
the total {\OIIIsemi} $\lambda$1665 detection confidence level is estimated to be $6.8\sigma$. The fluxes of the {\OIIIsemi} $\lambda$$1661$ and {\OIIIsemi} $\lambda$$1666$ emission lines are $0.92\times10^{-18}$ and $1.42\times10^{-18}$ $\mathrm{erg\ s^{-1}\ cm^{-2}}$, respectively. 
Here, the fluxes of  {\OIIIsemi} $\lambda$1661 and {\OIIIsemi} $\lambda$1666 lines are obtained by Gaussian function fitting.
We make a 1-dimensional pixel-noise spectrum based on the data of our LRIS 2-dimensional spectrum where the object is not included,
and estimate statistical errors of the line fluxes with the 1-dimensional pixel-noise spectrum
in the wavelength range same as the emission-line FWHM.

Although our LRIS slit covered three subcomponents of {\COSMOSno} \citep[see Figure 5 of ][]{Shibuya2014b}, 
there exist a non-negligible flux loss due to the large FWHM seeing sizes of $1.\!\!^{\prime\prime}0$--$1.\!\!\arcsec4$ whose median value is $\simeq1.\!\!\arcsec1$.
We estimate the slit loss correction factor, $F_{\rm LRIS}$, with the {\it Hubble Space Telescope} \citep[{\it HST;}][]{Skelton2014} $I_{814}$-band image whose point spread function (PSF) is convolved to match the median seeing size. 
The $F_{\rm LRIS}$ value is estimated to be 1.61. Applying the LRIS slit loss correction, we obtain $1.48\times10^{-18}$ and $2.29\times10^{-18}$ $\mathrm{erg\ s^{-1}\ cm^{-2}}$ for the {\OIIIsemi} $\lambda$$1661$ and {\OIIIsemi} $\lambda$$1666$ emission lines, respectively. Table \ref{tbl:c12805} summarizes the flux estimates of the {\OIIIsemi} emission lines with the $F_{\rm LRIS}$ correction.

\subsection{FMOS Data} \label{subsec:fmos}
Our FMOS observations with {\it J-}long and {\it H-}short bands were conducted over 2012 December 22--24 (UT) 
with seeing of $0.\!\!^{\prime\prime}8$ (FWHM).
The total exposure times were 2.0 and 6.5 hours for the {\it J-}long and {\it H-}short bands, respectively.
The spectral resolution of {\it J-}long and {\it H-}short bands are $R\simeq 1900$ and $2400$, which correspond to $158$ and $125$ km s$^{-1}$, respectively. Details of the FMOS observations and data reduction are given in K. Nakajima et al. in preparation. 

\begin{figure*}[t!]
\begin{center}
\includegraphics[scale=0.8, bb=0 0 610 229]{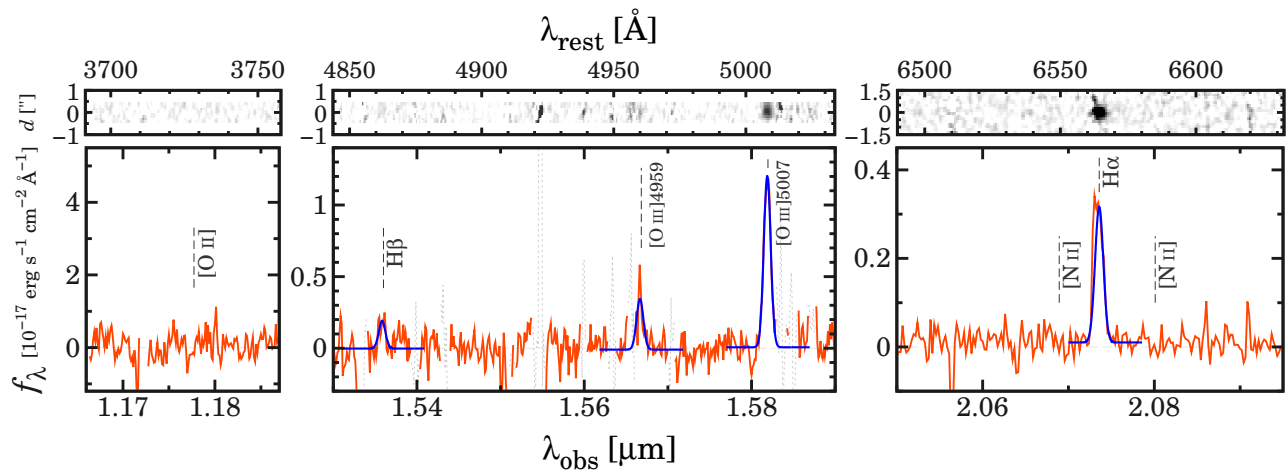}
\end{center}
\caption{{\footnotesize Rest-frame optical spectra of {\COSMOSno}. The left, center, and right panels present the FMOS {\it J}-long, {\it H}-short, and MOSFIRE {\it K}-band spectra, respectively. In each panel, the 2D and 1D spectra are shown at the top and the bottom, respectively. 
The blue solid lines denote the best-fit Gaussian functions for the detected emission lines.
 The gray vertical dashed lines indicate expected wavelengths of optical emission lines, {\OII}, {\Hb}, {\OIII}, {\NII}, and {\Ha} for the systemic redshift of $z_{\rm sys}=2.159$ (see Section \ref{subsec:fmos}). The gray dotted lines represent spectra affected by strong OH sky lines. }}
 \label{fig:spec_opt}
\end{figure*}

Figure \ref{fig:spec_opt} presents the FMOS spectra of {\COSMOSno}. 
In the {\it H-}short band spectrum, the {\OIII} $\lambda$$5007$ emission line 
is detected with an S/N of $\sim23$.
The {\Hb} and {\OIII} $\lambda$$4959$ emission lines are also detected with $S/N$ ratios of $\simeq 3.1$ and $4.5$, respectively.
The systemic redshift and the emission line fluxes are measured by simultaneous fitting with multiple Gaussian functions to the {\Hb} and {\OIII} $\lambda\lambda$$4959, 5007$ emission lines. 
We show the best fit Gaussian functions for {\Hb} and {\OIII} $\lambda$$4959, 5007$ emission lines in Figure \ref{fig:spec_opt}.
The flux errors are estimated in the same manner as Section \ref{subsec:lris}, but with
a 1-dimensional pixel-noise spectrum produced with the FMOS pipeline FIBRE-pac \citep[FMOS Image-Based Reduction Package;][]{Iwamuro2012}.
We obtain a systemic redshift of $z_{\rm sys}=2.159$ for {\COSMOSno}. The {\Hb}, {\OIII} $\lambda$$4959$, and {\OIII} $\lambda$$5007$ fluxes are $19.2\times10^{-18}$, $36.2\times10^{-18}$, and $124\times10^{-18}$ $\mathrm{erg\ s^{-1}\ cm^{-2}}$, respectively. In contrast, we find no emission line at the expected wavelength of {\OII} $\lambda$$3727$ in the {\it J-}long band spectrum. We place a $2\sigma$ upper limit on the {\OII} $\lambda$$3727$ line flux, $< 55.6\times10^{-18}$ $\mathrm{erg\ s^{-1}\ cm^{-2}}$, assuming that the intrinsic velocity dispersion is $135$ km s$^{-1}$ in FWHM. This intrinsic velocity dispersion is estimated with the detected {\OIII} $\lambda$$5007$ emission line.
 
In the observations for these optical emission lines, the FMOS fiber has covered only one of the three subcomponents in the {\COSMOSno} system. To obtain the total flux for all the three subcomponents, we estimate the flux loss correction factor, $F_{\rm FMOS}$, for the FMOS $1.\!\!\arcsec2$-diameter aperture fiber in the same manner as Section \ref{subsec:lris}. Using the $\!${\it HST} $I_{814}$-band image 
whose point-spread function is matched to FWHM=$0.\!\!^{\prime\prime}8$,
we measure the flux values for all the three subcomponents and the one covered by the FMOS fiber. 
The $F_{\rm FMOS}$ value is estimated to be 1.59. 
Because the morphology of {\COSMOSno} in the $I_{814}$ band
can be different from those of {\it J} and {\it H} bands for the FMOS data,
we make additional calculations for $F_{\rm FMOS}$ values 
with
the UltraVISTA \citep{McCracken2012} {\it J}- and {\it H}-band images 
in the $0.\!\!^{\prime\prime}8$ seeing size. 
The $F_{\rm FMOS}$ value estimated from the {\it J} band ({\it H} band) 
is 10\% larger (3\% smaller) than the one obtained from the $I_{814}$ band. Although 
these differences are small, we add these differences to the uncertainties of 
{\OII} $\lambda$$3727$, {\Hb}, and {\OIII}$\lambda\lambda$$4959, 5007$ line flux measurements.
Applying the FMOS fiber loss correction, we obtain $30.5\times10^{-18}$ for {\Hb}, $57.6\times10^{-18}$ for {\OIII} $\lambda$$4959$, $197\times10^{-18}$ for {\OIII} $\lambda$$5007$, and $<88.3\times10^{-18}$ $\mathrm{erg\ s^{-1}\ cm^{-2}}$ for {\OII} $\lambda$$3727$. Table \ref{tbl:c12805} summarizes the $F_{\rm FMOS}$-corrected fluxes of the {\Hb}, {\OIII}$\lambda\lambda$$4959, 5007$, and {\OII} $\lambda$$3727$ emission lines. 
 
\subsection{MOSFIRE Observations and Data} \label{subsec:mosfire}
We observed {\COSMOSno} with the Keck/MOSFIRE instrument on 2016 January 3 and 4.
We carried out $K$-band spectroscopy that covered the wavelength range of $1.95-2.4\ \mu$m, targeting the redshifted {\Ha} and {\NII} $\lambda$$6584$ emission lines.
We used the slit width of $0.\!\!\arcsec7$ and an ABAB dither pattern with individual exposures of $180\ \mathrm{seconds}$.
The total integration time was $2.4\ \mathrm{hours}$.
The sky conditions was clear throughout the observing run, with an average seeing size of FWHM$=0.\!\!\arcsec6$.
The pixel scale was 
$0\farcs18$, 
and the spectral resolution was $R=3620$ corresponding to $\Delta \lambda\sim6\ \mathrm{\AA}$ at $\lambda=2.162\ \mu$m.
We also took spectra of 
an A0V-type standard star,
HIP 56736, for our flux calibration.

We reduce these data using the MOSFIRE data reduction pipeline.\footnote{https://keck-datareductionpipelines.github.io/MosfireDRP/} 
This pipeline performs flat fielding, wavelength calibration, sky subtraction, and cosmic ray removal before producing a combined 
two-dimensional spectrum. 
We then extract one-dimensional spectra from the two-dimensional spectra using the {\sc iraf} task {\tt blkavg}. 
We sum the fluxes of a 2.$\!\!\arcsec$34 width (about a $5\times$ the average seeing) at each wavelength bin 
that covers more than 95\% fluxes of the point source along the slit direction.
The $3\sigma$ limiting flux density is $\simeq1\times10^{-18}\ \mathrm{erg\ s^{-1}\ cm^{-2}\ \AA^{-1}}$ 
in the wavelength range of $19500$--$23700\ \mathrm{\AA}$. 

The right panel in Figure \ref{fig:spec_opt} shows our MOSFIRE spectra at the wavelength of {\Ha} and {\NII} emission lines.
The {\Ha} emission line is detected 
with an S/N of $62$.
To estimate the total flux, we fit this {\Ha} line with a Gaussian function
(Figure \ref{fig:spec_opt}),
assuming the systemic redshift of $z_\mathrm{sys}=2.159$. 
This systemic redshift is determined by the emission lines in the FMOS spectra.
In the Gaussian fitting to the {\Ha} line, we do not use four spectral pixels in the shorter wavelength side, because these four pixels are contaminated by a strong night sky emission line (Figure \ref{fig:spec_opt}).
The statistical error of the {\Ha} line flux is estimated in the same manner as Section \ref{subsec:lris}.

Since {\COSMOSno} spatially extends in a direction perpendicular to the slit, we calculate a slit loss correction factor $F_{\rm MOSFIRE}$ to determine total line fluxes. 
We use the HST $I_\mathrm{814}$-band image, and convolve it with a Gaussian kernel to match its PSF to that of the MOSFIRE data.
We measure the total and slit-aperture fluxes of {\COSMOSno} on the image, and calculate a correction factor to be $F_{\rm MOSFIRE}=2.35$.
To evaluate the $F_{\rm MOSFIRE}$ uncertainty raised by
morphology differences between the $I_{814}$ and {\it Ks} bands,
we calculate $F_{\rm MOSFIRE}$ values 
with
UltraVISTA \citep{McCracken2012} {\it Ks}-band images 
under the assumption of the $0.\!\!^{\prime\prime}75$ seeing condition. The $F_{\rm MOSFIRE}$ value 
estimated from the {\it Ks}-band image is 5\% smaller than the one obtained from the $I_{814}$-band image. 
We include this difference of the slit loss correction factors to the flux uncertainties of
{\Ha} and {\NII} lines.

The total {\Ha} line flux, thus obtained, is $86.6\times10^{-18}\ \mathrm{erg\ s^{-1}\ cm^{-2}}$,
which is corrected for the slit loss.
Since an {\NII} emission line is not identified in our MOSFIRE spectra, we estimate a $2\sigma$ upper limit on the {\NII} line flux to be $3.06\times10^{-18}\ \mathrm{erg\ s^{-1}\ cm^{-2}}$ 
under the assumption of the intrinsic velocity dispersion of $140\ \mathrm{km\ s^{-1}}$ in FWHM. This intrinsic velocity dispersion is estimated with the {\Ha} emission line.
Table \ref{tbl:c12805} summarizes the flux estimates of the {\Ha} and {\NII} emission lines after the slit loss correction. 

We have thus obtained the LRIS, FMOS, and MOSFIRE data (Sections \ref{subsec:lris}--\ref{subsec:mosfire}) for {\COSMOSno}, which is one of our sample SFGs that are detailed in Section \ref{sec:sample}.

\section{SAMPLE} \label{sec:sample}
Our sample consists of a total of 41 galaxies, eleven SFGs at $z=1.4$--$3.6$ (Table \ref{tab:prop}) and the stack of 30 KBSS $z\sim 2$ galaxies 
taken from \citet{Steidel2016}. Hereafter, the stack of 30 KBSS galaxies is dubbed S16-stack,
which complements the eleven SFGs. 
We include the S16-stack in our sample, 
because \citet{Steidel2016} do not calculate the key quantities of N/O and {\q} by the direct $T_e$ method, but the strong line method.
Note that we refer to the redshift of the eleven SFGs as $z\sim 2$, 
because the median redshift of the sample galaxies is 2.2.
The {\OIIIsemi} $\lambda$$\lambda$1661, 1666 or {\OIII} $\lambda$4363 emission is detected in each spectrum of 
the eleven SFGs and the S16-stack (Table \ref{tab:forbidden_lines}). 
One of the eleven SFGs is {\COSMOSno} (Section \ref{sec:data}). The other 10 SFGs are taken form the literature: 7 gravitationally lensed galaxies \citep{Fosbury2003, Villar-Martin2004, Yuan2009, Christensen2012a, Christensen2012b, Bayliss2014, Pettini2010, James2014, Stark2014a} and 3 BX galaxies obtained by KBSS \citep{Steidel2014, Erb2016}. These 10 SFGs are found by our extensive search in the literature with the conditions of the availabilities of {\OIIIsemi} $\lambda\lambda$1661, 1666 / {\OIII} $\lambda$4363 and {\NII} $\lambda$6584 / {\NIIIsemi} $\lambda$1750, which are used for the $T_e$ method and the N/O determination, respectively. In this paper, we refer to the eleven galaxies by their IDs that are defined in Tables \ref{tab:prop}--\ref{tab:balmer_lines}.

The nebular line fluxes of the eleven galaxies and the S16-stack are summarized in Table \ref{tab:forbidden_lines} for forbidden and semi-forbidden lines and in Table \ref{tab:balmer_lines} for Balmer lines. The fluxes are normalized to H$\beta$ in Tables \ref{tab:forbidden_lines} and \ref{tab:balmer_lines}. 
Note that all the line fluxes in Tables \ref{tab:forbidden_lines} and \ref{tab:balmer_lines} are corrected for the telluric absorption, except for {\SMACS}
whose line fluxes corrected for telluric absorption are not available. 
For {\SMACS}, we estimate that the flux losses by the telluric absorption are 30\% for the H$\alpha$, H$\beta$, and {\NII} $\lambda$6584 lines, and 5\% for the {\OIII} $\lambda$5007 line 
based on Figure 2 of \citet{Christensen2012a}.
These flux losses are included in the errors of $T_e$(\textsc{O\,iii}), 12+$\log$(O/H)$_{T_e}$, $\log$(N/O)$_{T_e}$, and {\q} of {\SMACS}.
We correct the line fluxes for dust extinction with the Balmer decrement measurements and the extinction curve of \citet{Cardelli1989}, using the models of the case B recombination \citep[Table \ref{tbl:caseB}]{Osterbrock1989} with $n_e$= 100 cm$^{-3}$ and $T_e\sim 5000$--$20000$ K that we derive from the direct $T_e$ methods (Section \ref{subsec:Te}). We use the model values of Balmer decrements listed in Table \ref{tbl:caseB} for the three cases of $T_e=5000$, $10000$, and $20000$ K. We apply linear interpolations to obtain model Balmer decrements for our SFGs whose $T_e$ values are $5000$--$20000$ K.
We assume {\Ha}/{\Hb}=2.75, {\Hc}/{\Hb}=0.475, and {\Hd}/{\Hb}=0.26 for our SFGs with the measurement of $T_e>20000$ K. 
The electron temperature of ionized hydrogen, $T_e$(\textsc{H\,ii}) agrees well with $T_e$(\textsc{O\,iii}) in our photoionization model calculations with \textsc{Cloudy} \citep[version 13.03; ][]{Ferland1998, Ferland2013}, which are detailed in Section \ref{sec:analysis}.
We iteratively perform the dust correction so that the estimated $T_e$(\textsc{O\,iii}) match the initial assumption of $T_e$(\textsc{H\,ii}). 

We estimate the color excess $E(B-V)_{\rm neb}$ with {\Ha}/{\Hb}, except for {\SMACS}, {\SGAS}, and {\Lynx}.
The {\Ha} and {\Hb} lines of {\SMACS} are affected by the telluric absorption, and the {\Ha} line of {\SGAS} is not covered in the spectrum. 
Thus we use {\Hd}/{\Hc} and {\Hb}/{\Hc} to obtain $E(B-V)_{\rm neb}$ values of {\SMACS} and {\SGAS}, respectively.
We cannot calculate $E(B-V)_{\rm neb}$ of the {\Lynx}, because {\Lynx} only has {\Hb} emission. 

Table \ref{tab:prop} shows physical properties of the eleven galaxies and the S16-stack, which include
stellar masses $M_{\star}$ and star-formation rates (SFRs) taken from the literature.
The $M_{\star}$ values are obtained by spectral energy distribution (SED) fitting with stellar synthesis models, except for {\Lynx} and {\CSWA}. 
The $M_{\star}$ value of {\Lynx} is estimated with the instantaneous burst models required to reproduce the {\Hb} luminosity of 
{\Lynx} \citep{Villar-Martin2004}. The $M_{\star}$ value of {\CSWA} is missing, 
because there is no $M_{\star}$ measurement in the literature.
We choose the \citet{Chabrier2003} IMF in this paper. 
Because Chabrier or \citet{Salpeter1955} IMF is used to derive $M_{\star}$ in the literature, 
we subtract $0.20$ dex from $\log(M_{\star}/M_{\odot})$ of Salpeter IMF to obtain stellar masses of Chabrier IMF \citep{Madau2014}. 
$M_{\star}$ values of the our $z\sim2$ galaxies range in $\log(M_{\star}/M_{\odot})=7.49$--$10.57$. 
Their SFRs are $5.7$--$906$ $M_{\odot}\,{\rm yr}^{-1}$. 

In Table \ref{tab:prop}, we also present values of rest-frame {\Lya} equivalent width EW$_0$({\Lya}).
The six galaxies, {\COSMOS}, {\BXI}, {\BXII}, {\BXIII}, {\Lynx}, and {\AbellS}, show strong {\Lya} emission lines. 
These six galaxies indicate EW$_0$({\Lya}) $\gtrsim20$ {\AA} and are thus regarded as LAEs.

\input tab2.tex
\input tab3.tex
\input tab4.tex

\input tab5.tex


\subsection{AGN Activity} \label{subsec:AGN}
\begin{figure}
\begin{center}
\includegraphics[scale=0.4]{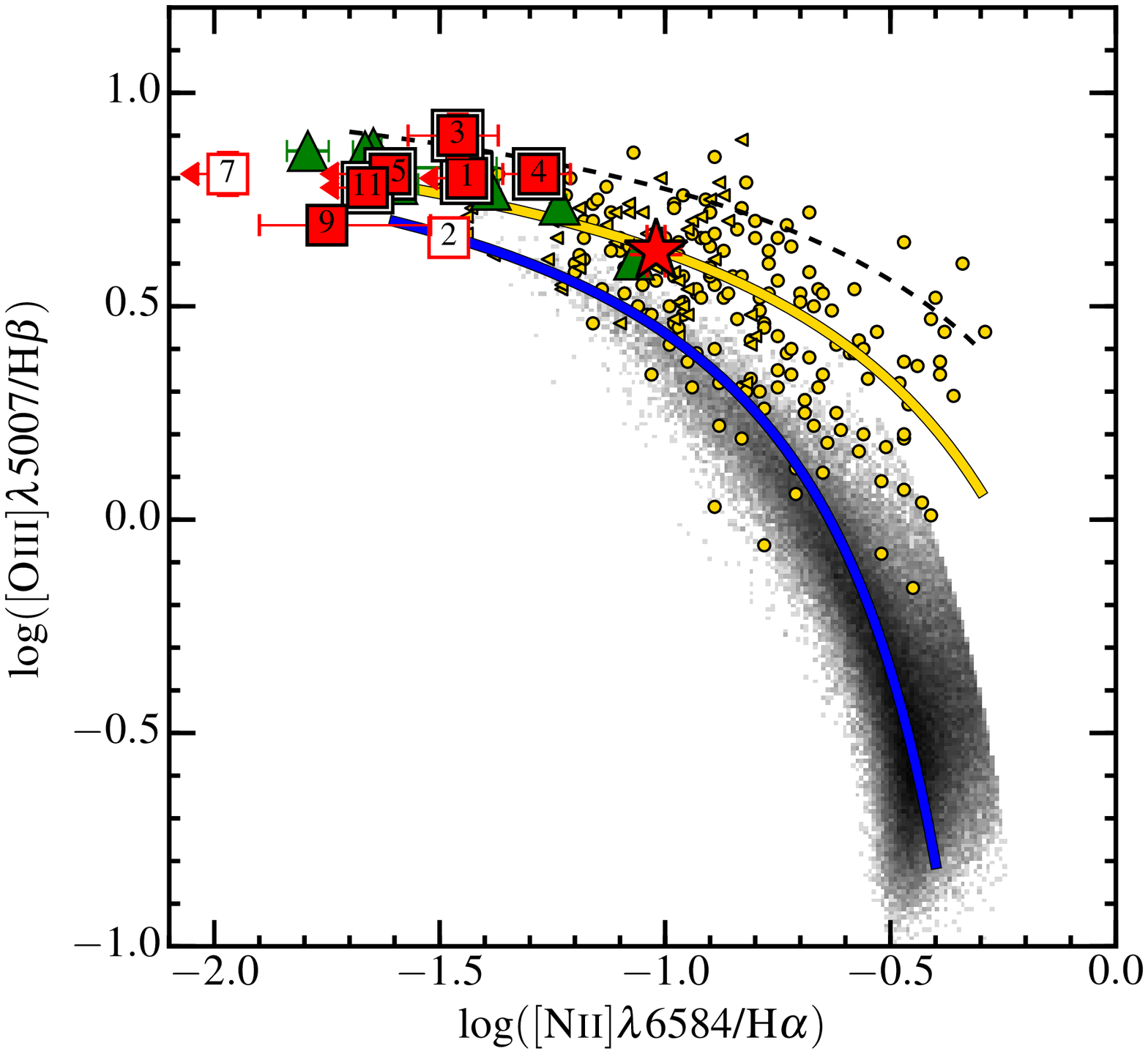}
\end{center}
\caption{{\OIII} 5007/{\Hb} vs. {\NII} 6584/{\Ha} BPT diagram for our $z\sim2$ galaxies. 
The red squares denote our galaxies, while the red filled squares are our galaxies in the reliable sample.
The numbers written on the red squares are the ID numbers of our galaxies (Table \ref{tab:prop}--\ref{tab:balmer_lines}). In our $z\sim2$ galaxies, 
only 8 galaxies are plotted because the spectra of the other 3 galaxies, {\Lynx}, {\SGAS}, and {\AbellS}, do not cover the wavelength of the {\NII} line. 
The yellow circles are the UV-selected galaxies at $z=2.3$ from KBSS \citep{Steidel2014}. 
The yellow triangles are the same as the yellow circles,
but for the KBSS galaxies with the upper limits of {\NII}/{\Ha}. 
The yellow solid line is the best-fit BPT sequence for the KBSS galaxies. 
The grayscale indicates the distribution of the local SDSS galaxies. 
The blue solid line is the best-fit sequence of the SDSS galaxies \citep{Kewley2008}. 
The black dashed line represents the maximum theoretical model of star forming galaxies at $z=2$ \citep{Kewley2013a}. 
We also plot the S16-stack with the red star, and nine green pea galaxies with the green triangles \citep[][see Section \ref{subsec:FMR}]{Amorin2012, Jaskot2013}.
The $z\sim2$ galaxies with EW$_0$({\Lya}) $>20$ {\AA} are marked with the large black squares.
For our galaxies, the upper limits of {\NII}/{\Ha} correspond to the $2\sigma$ levels.
\label{fig:bpt}}
\end{figure}

Figure \ref{fig:bpt} presents our galaxies on the BPT diagram. 
Because the {\NII} emission lines are not observed in 3 galaxies of {\Lynx}, {\SGAS}, and {\MACS}, we only plot the other 8 galaxies here. 
These 8 galaxies are located in or near the SFG region that is defined by the models of starbursts at $z=2$ \citep{Kewley2013a}. 
Although {\BXI} is located on the edge of the SFG and AGN regions,
the rest-frame UV and optical spectra of {\BXI} show no signatures of an AGN \citep{Steidel2014}. 
We thus regard these 8 galaxies as SFGs with no significant AGN activity. In Figure \ref{fig:bpt}, the {\NII}/{\Ha} upper limit of {\AbellYK} is very small. This is probably because the {\NII} line estimation would include a systematic uncertainty raised by the low throughput spectrum at the wavelength of {\NII} \citep{Yuan2009}. 
We find that, at fixed values of {\NII}/{\Ha}, the majority of our galaxies have the {\OIII}/{\Hb} values that are higher than the $z\sim 0$ SDSS SFG sequence. 
This trend of the high {\OIII}/{\Hb} values is consistent with those found in previous studies of galaxies at $z\sim2$ \citep{Shapley2005b, Erb2006a, Liu2008, Steidel2014, Shapley2015}.

We then investigate AGN activities of the 3 galaxies, the {\Lynx}, {\SGAS}, and {\MACS}, which cannot be plotted in the BPT diagram (Figure \ref{fig:bpt}).
{\SGAS} has a spatially extended emission region and no high ionization lines, such as \textsc{N\,v}\,$\lambda$1240 and [Ne \textsc{v}] $\lambda\lambda$3346, 3426 \citep{Bayliss2014}. Moreover, {\SGAS} falls on the SFG region of {\OII} $\lambda$3727, {\NeIII} $\lambda$3869 diagnostic diagram
obtained by \citet{Perez-Montero2007}. Thus, we regard {\SGAS} as an SFG.
We also classify {\MACS} as an SFG because the high ionization lines of \textsc{N\,iv}]\,$\lambda$1487, \textsc{C\,iv}\,$\lambda$1548,1550, and He\,\textsc{ii}\,$\lambda$1640 are very weak \citep{Stark2014a}. 
Although {\Lynx} does not show the high ionization line of \textsc{N\,v}\,$\lambda$1240, 
the large values of 
\textsc{N\,iv}]\,$\lambda\lambda$1483,1487/\textsc{N\,iii}]\,$\lambda$1750 = 2.3 and \textsc{C\,iv}\,$\lambda\lambda$1548,1551/\textsc{C\,iii}]\,$\lambda\lambda$1907,1909 = 6.1
can be produced by a hard ionizing source spectrum. 
The spectrum of {\Lynx} may be contaminated by an AGN activity.
In summary, we classify {\SGAS} and {\MACS} as SFGs, 
while {\Lynx} could possibly be an AGN.

\subsection{Reliable Sample} \label{subsec:reliability}
We aim at studying SFGs at $z\sim 2$ with no AGN contamination and no systematic uncertainties.
We identify that four out of our eleven galaxies, {\SMACS}, {\Lynx}, {\AbellYK}, and {\SGAS}, 
would have potential problems of AGN and/or systematics.
{\SMACS} has line fluxes of {\Hb}, {\OIII}$\lambda\lambda$$4959, 5007$, {\Ha}, and {\NII} $\lambda$6584
that are under the potentially large influences of telluric absorption lines \citep{Christensen2012a}.
{\Lynx} has only one Balmer line (\Hb) that is detected. No Balmer decrement analysis can be applied to {\Lynx} for dust-extinction correction. 
Although the SED fitting suggests that {\Lynx} is free from dust extinction \citep{Fosbury2003, Villar-Martin2004}, the dust-extinction corrected line fluxes of {\Lynx} are not as reliable as those of the other galaxies. Moreover, {\Lynx} may be contaminated by an AGN (Section \ref{subsec:AGN}). 
Similarly, {\AbellYK} has a potential uncertainty in the dust-extinction estimates. 
The emission line of {\Ha} is detected in the wavelength with the low instrumental response that may include an unknown systematics.
As presented in Section \ref{sec:analysis}, the low values of 12+log(O/H) and $q_{\rm ion}$ of {\AbellYK} would be produced by 
the large potential uncertainty of the {\Ha} line flux that gives the uncertainty in the $E(B-V)_{\rm neb}$ estimation.
Another uncertainty in the dust extinction correction may exist in {\SGAS}.
The measurement of {\Hc} for {\SGAS} would not be reliable due to the poor atmospheric throughput at the wavelength of {\Hc}.
Although we obtain $E(B-V)_{\rm neb}$=0.0 for {\SGAS} from the Balmer decrement measurement (Table \ref{tab:balmer_lines}), 
this value is not consistent with the extinction of $A_{\rm v}=1.0$ estimated from the SED fitting \citep{Bayliss2014}.
We find that there are no potential problems in the other 7 galaxies ({\COSMOS}, {\BXI}, {\BXII}, {\BXIII} {\CSWA}, {\MACS}, and {\AbellS}) and the S16-stack.
We make a subsample of these 7 galaxies and the S16-stack, and refer the subsample as ``reliable sample''.

\section{ANALYSIS} \label{sec:analysis}
In this section, we estimate electron temperatures ($T_e$), oxygen abundances (O/H), ionization parameters ({\q}), and nitrogen to oxygen abundance ratios (N/O) for the eleven galaxies and the S16-stack in our sample. 
In this paper, we regard these estimates as the average values of the galaxies, because the current telescopes and 
instruments cannot resolve the {\HII} regions in $z\sim2$ galaxies.
The uncertainties of $T_e$, O/H, {\q}, and N/O are estimated by the Monte Carlo simulations, where the uncertainties in the flux measurements are propagated to the $T_e$, O/H, {\q}, and N/O estimates. 
We also include the uncertainties of the dust extinction correction.
The uncertainties obtained in this way are presented in Table \ref{tab:prop}.

We construct plane-parallel photoionization models with \textsc{Cloudy}. We assume solar abundance ratios for all of the elements, except for helium, carbon, and nitrogen. 
We use forms of \citet{Pilyugin2012} and \citet{Dopita2006b} for nitrogen and carbon+helium, respectively.
Photoionization models are calculated with the gas-phase metallicities of 12+$\log$(O/H)= (7.54, 7.69, 7.99, 8.17, 8.29, 8.39, 8.59, 8.69), 
the ionization parameters of $\log$({\q} /cm s$^{-1}$) = (7.0, 7.2, 7.4, 7.6, 7.8, 8.0, 8.2, 8.4, 8.6, 8.8, 9.0), and 
hydrogen density of n$_{\rm H}$ = 250 cm$^{-3}$ given by \citet{Steidel2014}, \citet{Shimakawa2015}, and \citet{Sanders2016a}.
Under the assumption that the hydrogen atoms are fully ionized in the {\HII} regions, the hydrogen density of $n_{\rm H}=$250 cm$^{-3}$ is justified for our sample, because the median value of $n_e\sim300$ cm$^{-3}$ are obtained for five galaxies of ID-2--5 and ID-9 that have $n_e$ indicators of {\OII} $\lambda\lambda$3727, 3729 flux measurements \citep{Christensen2012b, Steidel2014, James2014}.
The incident radiation spectra are generated by stellar synthesis modeling with {\sc Starburst99} \citep{Leitherer1999, Leitherer2010, Leitherer2014}. 
In the {\sc Starburst99} models, we adopt instantaneous burst models at zero age with a \citet{Kroupa2001} IMF 
and lower and upper mass limits of 0.1 and 120 $M_{\odot}$, respectively. 
The stellar atmosphere models of WM-Basic \citep{Pauldrach2001} and CMFGEN \citep{Hillier1998, Hillier1999} are used here.
Note that there remains an uncertainty in the selection of stellar atmosphere models, which can affect the calculation results of the \textsc{Cloudy} models.
Stellar metallicities are matched to the gas-phase ones.
Our calculations are terminated when the hydrogen ionization fractions reach $\le 1$\%.

\subsection{Electron Temperature} \label{subsec:Te}

\begin{figure}
\begin{center}
\includegraphics[scale=0.4]{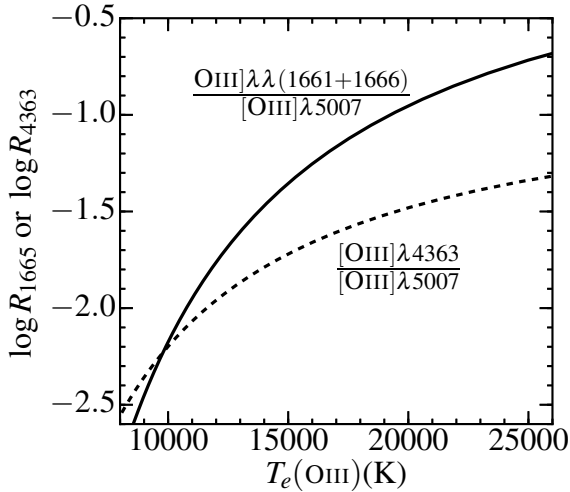}
\end{center}
\caption{Flux ratios, $R_{1665}$ (solid line) and $R_{4363}$ (dashed line), as a function of electron temperature $T_e$(\textsc{Oiii}).
The flux ratios are defined as $R_{1665}=${\OIIIsemi} $\lambda \lambda$1661,1666/{\OIII} $\lambda$5007 and $R_{4363}=${\OIII} $\lambda$4363/{\OIII} $\lambda$5007. 
These $R_{1665}$ and $R_{4363}$ values are obtained by the \textit{nebular.temden} routine of \textsc{iraf/stsdas} under the assumption of the low electron density limit. 
\label{fig:temp}}
\end{figure}

\begin{figure}
\begin{center}
\includegraphics[scale=0.56]{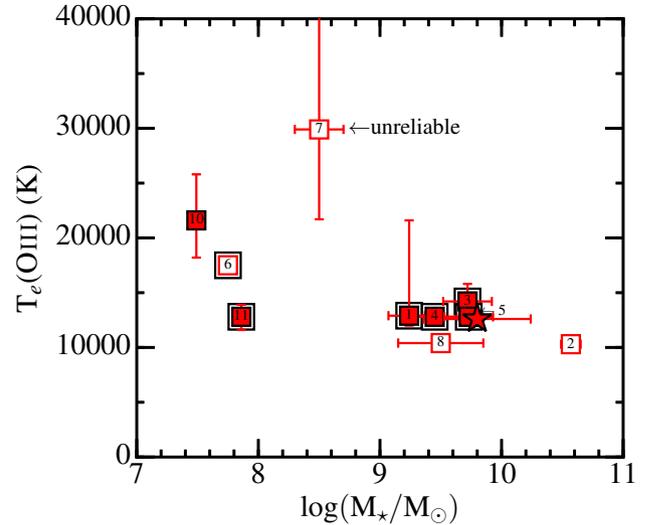}
\end{center}
\caption{Electron temperature measurements of our $z\sim2$ galaxies as a function of stellar mass. 
The red squares denote our galaxies, while the red filled squares are our galaxies in the reliable sample.
The numbers written on the red squares are the ID numbers of our galaxies (Tables \ref{tab:prop}--\ref{tab:balmer_lines}). 
The data point of {\CSWA} is not shown because an $M_{\star}$ measurement of {\CSWA} is not available.
We also plot the S16-stack with the red star.
The $z\sim2$ galaxies with EW$_0$({\Lya}) $>20$ {\AA} are marked with the large black squares.
\label{fig:temp_mass}}
\end{figure}

We estimate $T_e$ values of {\HII} regions in our galaxies with the (semi-)forbidden lines, {\OIIIsemi} $\lambda \lambda$1661,1666, {\OIII} $\lambda$4363, and {\OIII} $\lambda$5007. Figure \ref{fig:temp} presents the $T_e$ dependence of the two flux ratios, $R_{1665}\equiv${\OIIIsemi} $\lambda \lambda$(1661+1666)/{\OIII} $\lambda$5007 (solid line) and $R_{4363}\equiv$ {\OIII} $\lambda$4363/{\OIII} $\lambda$5007 (dashed line). From the measurements of $R_{1665}$ and $R_{4363}$, we obtain $T_e$ values with the \textit{nebular.temden} routine \citep{Shaw1995} in \textsc{iraf/stsdas}, assuming the low $n_e$ limit. 
If $n_e$ is significantly lower than the critical densities, these (semi-)forbidden line fluxes only depend on $T_e$. 
The critical densities of {\OIII} $\lambda$5007, {\OIII} $\lambda$4363, and {\OIIIsemi} $\lambda \lambda$1661,1666 are $6.4\times10^5$, $2.8\times10^7$, and $3.4\times10^{10}$ cm$^{-3}$, respectively.
It is reported that the $n_e$ values of $z=2-3$ SFGs are clearly low, $\lesssim10^{4}$ cm$^{-3}$ based on the estimates with the line ratios of {\OII} $\lambda\lambda$3727/3729 and {\SII} $\lambda\lambda$6716/6731 \citep{Shimakawa2015, Sanders2016a}. 
Thus, the dependence of $n_e$ is negligible in our $T_e$ estimates.
Because one of the {\OIIIsemi} doublet lines, {\OIIIsemi} $\lambda1661$, is not detected in our 3 galaxies ({\CSWA}, {\MACS}, and {\AbellS}), 
we assume {\OIIIsemi} $\lambda\lambda1666/1661=2.46$
for these 3 galaxies. It should be noted that the ratio of {\OIIIsemi} $\lambda\lambda1666/1661=2.46$ marginally 
depends either on $T_e$ or $n_e$. We confirm the marginal dependence on $T_e$ or $n_e$ with our \textsc{Cloudy} model calculations, suggesting that a change of the {\OIIIsemi} ratio is less than 0.01 dex in the parameter ranges of log($q_{\rm ion}$/cm s$^{-1}$)$=7.5$--$9.0$ and 12+$\log$(O/H)$=7.54$--$8.69$.
In this way, we obtain $T_e$ values of our galaxies, which are typically $\sim 10,000$--$20,000$ K.
Table \ref{tab:prop} and Figure \ref{fig:temp_mass} summarize the $T_e$ values of our galaxies.

Note that {\AbellYK} and {\MACS} have high $T_e$ values of $\sim$30,000 K and $\sim$22,000 K, respectively. 
The unrealistically high $T_e$ value of {\AbellYK} is due to the large {\Ha}/{\Hb} ratio of $5.03$, which is probably biased by the systematic error of the telluric absorption correction for the {\Ha} line flux estimate.
Thus we do not include {\AbellYK} in the reliable sample.
In contrast with {\AbellYK}, we find no possible systematics in the {\MACS} flux measurements. 
However, it is unclear whether we can apply \textit{nebular.temden} over $20,000$ K because \textit{nebular.temden} uses the {\OIII} collisional strengths that are confirmed to be valid only below $20,000$ K \citep{Seaton1975}. Thus we compare the $T_e$ values that are obtained with the \textit{nebular.temden} routine and the \textsc{Cloudy} model.  \textsc{Cloudy} uses the {\OIII} collisional strengths that are valid from $\sim1,000$ K to $\sim100,000$ K \citep{Dere1997}.
We find that the $T_e$ differences between \textit{nebular.temden} and \textsc{Cloudy} are less than 1 per cent around $22,000$ K.
\footnote{We also confirm that the $T_e$ differences between \textit{nebular.temden} and \textsc{Cloudy} are less than 1 per cent in the range of $10,000$--$22,000$ K.} 
Thus we think that the $T_e$ value of {\MACS} is reliably estimated.

In addition, we find that the $T_e$ value of {\MACS} ($\sim$22,000 K) falls in the temperature range of the existing {\HII} regions \citep{Izotov2012}. This indicates that the $T_e$ value of {\MACS} ($\sim$22,000 K) is realistic.

\subsection{Oxygen Abundance} \label{subsec:oxygen}
We estimate oxygen abundances with the direct $T_e$ methods. Here, we adopt
\begin{equation}
	\frac{\textrm{O}}{\textrm{H}} = \frac{\textrm{O}^{+}+\textrm{O}^{2+}}{\textrm{H}^{+}}
\end{equation}
for {\HII} regions of SFGs. We ignore oxygen ions of O$^{3+}$ 
because the ionization potential of O$^{3+}$ 
is $54.9$ eV, too high for the stellar radiation to produce O$^{3+}$. 
We estimate ionic oxygen abundances of $\textrm{O}^{+}$ and $\textrm{O}^{2+}$ with the relations \citep{Izotov2006},
\begin{eqnarray}
	12+\log\left(\frac{\textrm{O}^{+}}{\textrm{H}^{+}}\right)& = &\log\left(\frac{\textrm{\OII}\,\lambda3727}{\textrm{H}\beta}\right) +5.961 \nonumber \\
	&& + \frac{1.676}{t_2} -0.40 \log t_2 - 0.034t_2 \label{eq:ophp}
\end{eqnarray}
\begin{eqnarray}
	12+\log\left(\frac{\textrm{O}^{2+}}{\textrm{H}^{+}}\right) &=& \log\left(\frac{\textrm{\OIII}\,\lambda\lambda4959+5007}{\textrm{H}\beta}\right)+6.200 \nonumber \\
	&& + \frac{1.251}{t_3} - 0.55 \log t_3 - 0.014t_3, \label{eq:opphp}
\end{eqnarray}
where we omit the term of $n_e$ that is negligibly small.
In Equations (\ref{eq:ophp}) and (\ref{eq:opphp}), we define $t_2=10^{-4}T_e$(\textsc{O\,ii}) and $t_3=10^{-4}T_e$(\textsc{O\,iii}),
where $T_e$(\textsc{O\,ii}) and $T_e$(\textsc{O\,iii}) are electron temperatures measured with [\textsc{O\,ii}] and [\textsc{O\,iii}], respectively.
Because auroral lines of [\textsc{O\,ii}] ({\OII} $\lambda \lambda$7320, 7330) are not observed in our galaxies, 
we assume
\begin{equation}
	t_2 = 0.7 t_3+ 0.3 \label{eq:t2}
\end{equation}
\citep{Campbell1986, Garnett1992}. 
Note that our conclusions of the oxygen abundance little depend on the assumption of $t_2$.
This is because our galaxies generally have high values of log(O$^{2+}$/O$^+$)$\sim 0.8$--$1.0$ and 
the contribution of O$^{2+}$/H$^+$ to the oxygen abundance estimate is much larger than the one of O$^{+}$/H$^+$.
In fact, if we change $T_e$(\textsc{O\,ii}) by $\pm$2000 K, 
the oxygen abundance 12+log(O/H) differs only by $<0.04$ dex.
By these calculations, we obtain 12+log(O/H)=$7.3$--$8.4$ for our galaxies.
Our 12+log(O/H) estimates are summarized in Table \ref{tab:prop}.

\subsection{Ionization Parameter} \label{subsec:q}
We estimate the ionization parameters {\q} with O$_{32}$ using the grids of the photoionization models \citep{Kewley2002}. 
Here, {\q} is defined as 
\begin{equation}
	q_{\rm ion} = \frac{S_{H^0}}{n_{H}},
\end{equation}
where $n_{H}$ is the density of hydrogen
and $S_{H^0}$ is the flux of ionizing photons at the inner surface of the plane-parallel slab. 
Although an O$_{32}$ measurement is sensitive to both {\q} and 12+log(O/H),
we have the estimate of 12+log(O/H) given by the direct $T_e$ methods (Section \ref{subsec:oxygen}).
Thus, our {\q} measurements do not include systematic uncertainties 
originated from the degeneracy with 12+log(O/H) estimates that is found
in the strong line method results.
Table \ref{tab:prop} summarizes the {\q} estimates.
Seven out of our eleven galaxies have the values of $\log$($q_{\rm ion}$/cm s$^{-1}$)$=7.7-8.6$, 
while the other four galaxies have the lower limits at $\log$($q_{\rm ion}$/cm s$^{-1}$)=$7.8-8.6$.
The S16-stack have $\log$($q_{\rm ion}$/cm s$^{-1}$)$=$7.70.

\subsection{Nitrogen-to-Oxygen Abundance Ratio} \label{subsec:N/O}

\begin{figure}
\begin{center}
\includegraphics[scale=0.4]{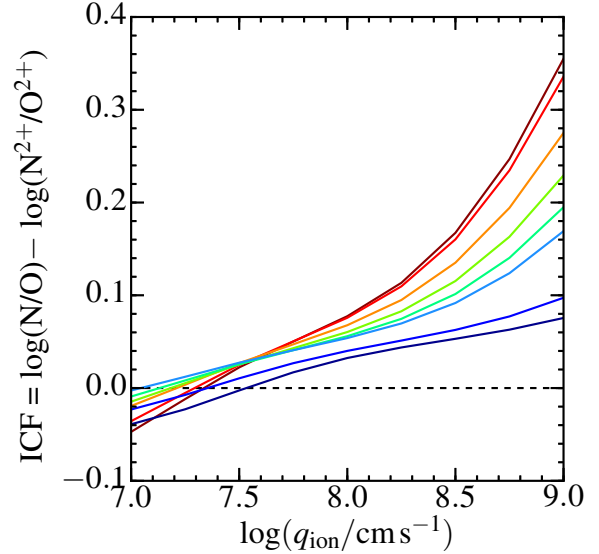}
\end{center}
\caption{Ionization correction factors (ICFs) obtained from our \textsc{Cloudy} models. The solid lines are ICFs with the color codes
defined by 12+log(O/H). The 12+log(O/H) values are 7.54, 7.69, 7.99, 8.17, 8.29, 8.39, 8.59 and 8.69, which correspond to
the colors from dark red to dark blue. The black dashed line indicates ICF=0.
\label{fig:icf}}
\end{figure}

We calculate N/O values with the line fluxes of {\NII} $\lambda$6584 and {\NIIIsemi} $\lambda$1750 
by the direct $T_e$ methods. 

For the six $z\sim2$ galaxies ({\SMACS}, {\BXI}, {\BXII}, {\BXIII}, {\AbellYK}, and {\CSWA}) and the S16-stack with {\NII} $\lambda$6584 and {\OII} $\lambda$3727 lines, 
we calculate
\begin{eqnarray}
	\log\left(\frac{\textrm{N}^{+}}{\textrm{O}^{+}}\right) &=& \log\left(\frac{\textrm{\NII}\, \lambda6584}{\textrm{\OII}\,\lambda3727} \right) + 0.400 \nonumber \\
	&& - \frac{0.726}{t_2} -0.02 \log t_2 + 0.007t_2 \label{eq:npop},
\end{eqnarray}
which is given by \citet{Izotov2006} under the assumption of {\NII} $\lambda\lambda6584/6548 = 2.95$. 
We estimate N/O with
\begin{equation}
	\frac{\textrm{N}}{\textrm{O}} \simeq \frac{\textrm{N}^{+}}{\textrm{O}^{+}} \label{eq:no1}
\end{equation}
\citep{Peimbert1967}. Note that
the ionic ratio of N$^{+}/$O$^{+}$ is a good indicator of N/O because ionization potentials of neutral oxygen and nitrogen are almost the same. 
Our \textsc{Cloudy} models confirm that the uncertainties of the N/O estimates from Equations (\ref{eq:npop}) and (\ref{eq:no1}) are less than 0.04 dex and 0.05 dex, respectively, in the parameter ranges of log(\q/cm\,s$^{-1}$)$=7.0$--$9.0$ and 12+$\log$(O/H)$=7.54$--$8.69$
(see also \citealt{Garnett1990}).


For the other 5 galaxies to which we cannot apply Equation (\ref{eq:npop}), 
we use the fluxes of {\NIIIsemi} $\lambda$1750 and ${\textrm{\OIIIsemi}\,\lambda\lambda1661+1666}$ to calculate
\begin{eqnarray}
	\log\left(\frac{\textrm{N}^{2+}}{\textrm{O}^{2+}}\right) &=& \log\left(\frac{\textrm{\NIIIsemi}\,\lambda 1750}{\textrm{\OIIIsemi}\,\lambda\lambda1661+1666}\right) \nonumber \\ 
	&& -0.674 - \frac{0.187}{t_3} \label{eq:nppopp}
\end{eqnarray}
\citep{Garnett1995}. 
We then estimate the N/O from the N$^{2+}$/O$^{2+}$ ratio with the ionization correction factor (ICF), 
\begin{equation}
	\log \left( \frac{\textrm{N}}{\textrm{O}} \right) = \log \left( \frac{\textrm{N}^{2+}}{\textrm{O}^{2+}} \right) + \textrm{ICF}. \label{eq:no2}
\end{equation}
We obtain the ICF values from our \textsc{Cloudy} models, which are shown in Figure \ref{fig:icf}.
Although the \q\ dependence of ICF is not very large ($<0.2$ dex in the parameter ranges of 
log(\q/cm\,s$^{-1}$)$=7.5$--$8.6$ and 12+$\log$(O/H)$=7.54$--$8.69$ in Figure \ref{fig:icf}), we use the
correction factors of ICF.

By these calculations, we find that six out of our eleven galaxies have values of $\log$(N/O)$_{T_e}=(-1.6)$--$(-1.0)$, 
while the other five galaxies have upper limits ranging from $-1.7$ to $-1.3$. We obtain the value of $\log$(N/O)$_{T_e}$=$-1.21$ for the S16-stack. Table \ref{tab:prop} presents the N/O values of our sample.

Note that it is worth comparing the N/O estimates from 
N$^{+}$/O$^{+}$ (eqs. \ref{eq:npop} and \ref{eq:no1})
and N$^{2+}$/O$^{2+}$ (eqs. \ref{eq:nppopp} and \ref{eq:no2}) ratios of an object
for the consistency check.  However, we cannot make
this comparison, because there are no objects with both N$^{+}$/O$^{+}$ and N$^{2+}$/O$^{2+}$
measurements (or constraints). Although ID-1 and 10 have the N/O upper limits estimated from N$^{2+}$/O$^{2+}$
ratios, these two objects have no constraints on N/O obtained by N$^{+}$/O$^{+}$ ratios, due to the
given upper limits of both {\OII} $\lambda$3727 and {\NII} $\lambda$6584 fluxes.

　
\section{RESULTS AND DISCUSSION} \label{sec:result}
In this section, we present the results of our analysis given in Section \ref{sec:analysis}, 
and compare them with average measurements from stacks of the 208,529 SDSS galaxies given by \citet{Andrews2013} whose physical properties are determined by the direct $T_e$ methods. 
We refer to these SDSS galaxy stacks as ``local stacks''. 
\citet{Andrews2013} made two kinds of the local stacks, $M_{\star}$ and $M_{\star}$--SFR stacks. 
The $M_{\star}$ stacks are the stacked spectra binned in $M_{\star}$, while the $M_{\star}$--SFR stacks are those binned in $M_{\star}$ and SFR. 
The SFR and $M_{\star}$ values are taken from the MPA-JHU value-added catalog\footnote{http://wwwmpa.mpa-garching.mpg.de/SDSS/DR7/} \citep{Kauffmann2003b, Brinchmann2004, Salim2007}. 
The $M_{\star}$ values are adjusted to the Chabrier IMF for comparison.
We use N/O values determined by \citet{Andrews2013}, whereas we estimate {\q} of the local stacks in the same manner as Section \ref{subsec:q}.

\begin{figure}
\begin{center}
\includegraphics[scale=0.33]{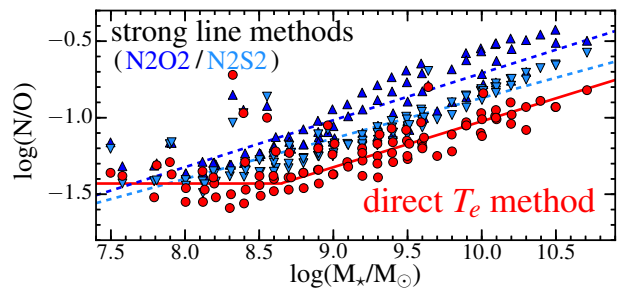}
\end{center}
\caption{
$\log$(N/O) as a function of $\log(M_\star/M_\odot)$ for the $z\sim 0$ galaxies
used in our comparisons. The red circles indicate the SDSS galaxies whose
$\log$(N/O) is derived with the direct $T_e$ methods \citep{Andrews2013}. 
The blue triangles and cyan inverted triangles are the
same as the red circles, but with $\log$(N/O) values estimated by the strong line methods of N2O2 and N2S2, respectively, 
calibrated by \citet{Perez-Montero2009} (see text).
The red-solid and blue-dashed lines denote the linear functions representing
the averages of the red circles and blue triangles, respectively, 
with the statistical weights of the galaxy numbers. This plot demonstrates that
the strong line method gives a systematic bias that increases the $\log$(N/O) estimates,
and that the direct $T_e$ methods are needed for the fair comparisons of the $z\sim 0$ and $z\sim 2$
galaxies.
\label{fig:no_m_hikaku}}
\end{figure}

In Figure \ref{fig:no_m_hikaku}, we plot $\log$(N/O) as a function of $\log(M_\star/M_\odot)$ for the $z\sim 0$ SDSS galaxies. 
Figure \ref{fig:no_m_hikaku} shows not only the $\log$(N/O) estimates from the direct $T_e$ methods, but also those from 
the strong line methods calculated with {\NII} $\lambda$6584/{\OII} $\lambda$3727 ($\equiv$N2O2) and {\NII} $\lambda$6584/{\SII} $\lambda\lambda$(6717+6731) ($\equiv$N2S2) by the procedures of \citet{Perez-Montero2009}.
This plot shows that the strong line method gives a systematic bias that elevates the $\log$(N/O) estimates.
This systematic bias may be made by the incorrect calibration 
due to the missing low $T_e$ systems.
Thus, the direct $T_e$ methods are needed for the fair comparisons of the $z\sim 0$ and $z\sim 2$
galaxies. We use $\log$(N/O) estimates from the direct $T_e$ methods both for the 
$z\sim 0$ and $2$ galaxies. 

As explained in Section \ref{sec:sample}, we assume the dust extinction curve of \citet{Cardelli1989} in this paper.
However, the assumption of the dust extinction curve might change the results of this paper.
Thus we also calculate the values of $T_e$, O/H, {\q}, and N/O under the assumption of the dust extinction curve of \citet{Calzetti2000}, and check whether the results in this section change.
We have confirmed that the results in this section do not change by the assumption of the dust extinction curve. 
This is because our galaxies have low $E(B-V)_{\rm neb}$ values (Table \ref{tab:balmer_lines}).

\subsection{Mass--Metallicity--SFR Relation} \label{subsec:FMR} 
\begin{figure}
\begin{center}
\includegraphics[scale=0.40]{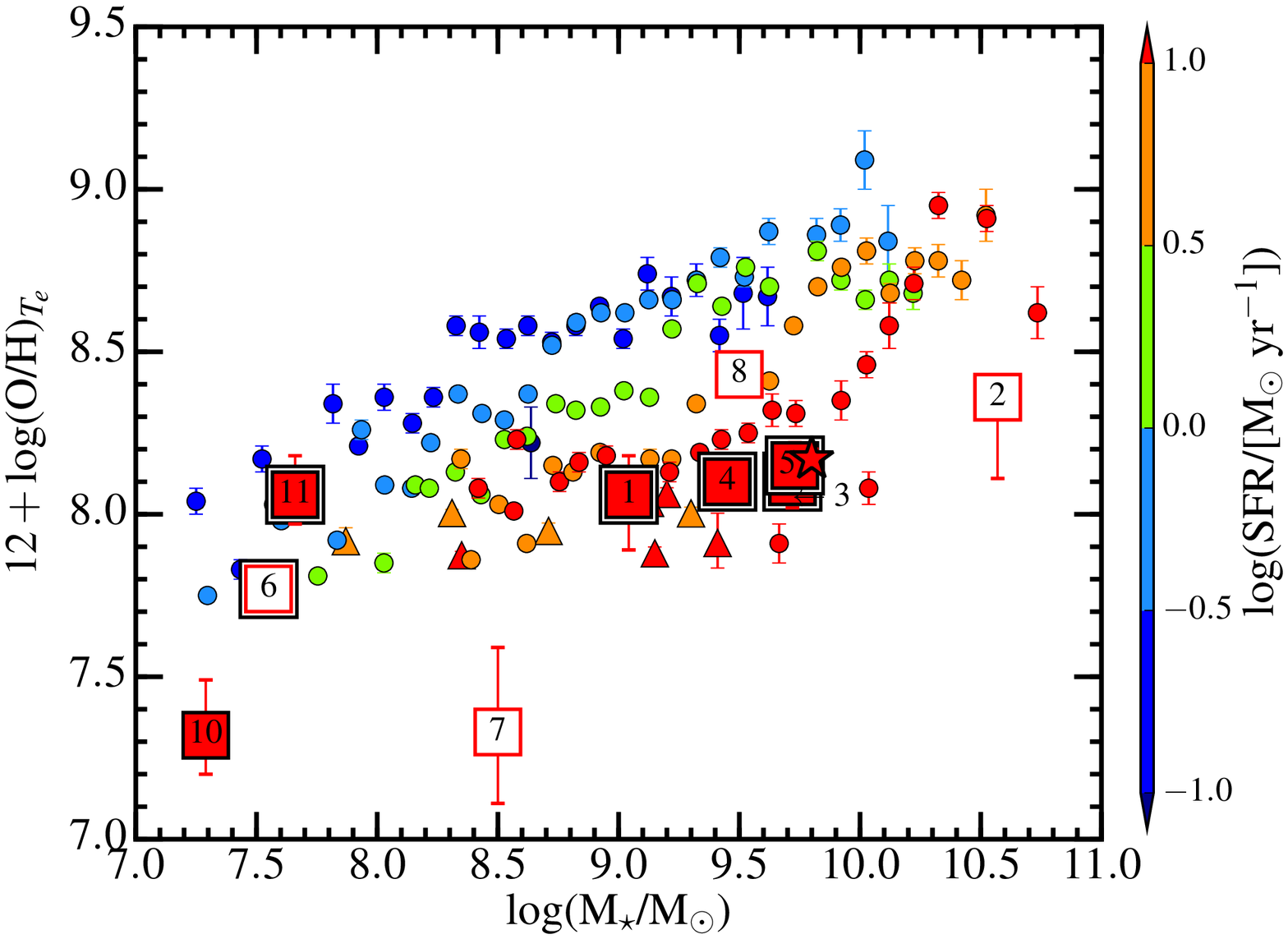}
\includegraphics[scale=0.40]{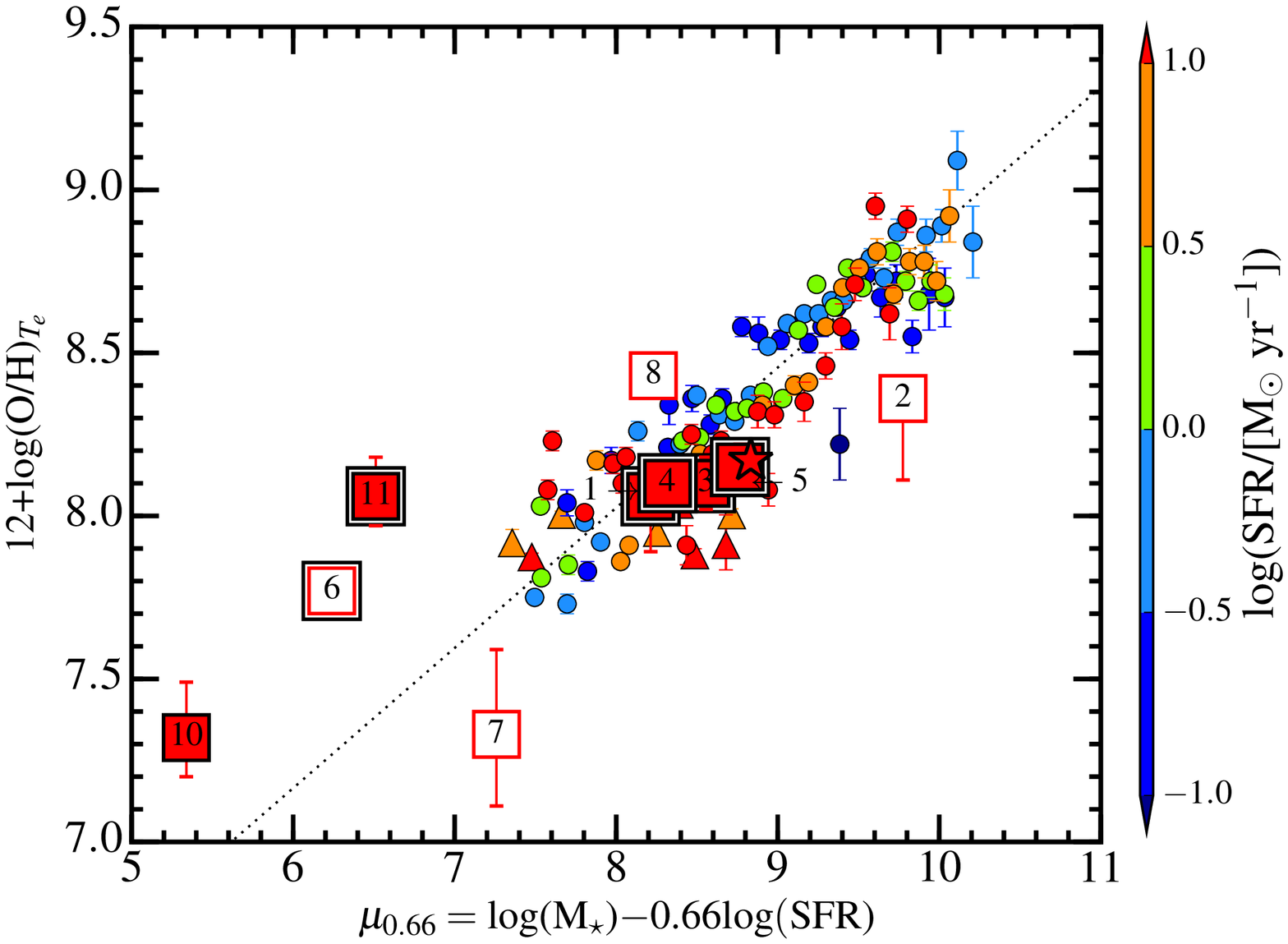}
\end{center}
\caption{$M_{\star}$ vs. 12+log(O/H)$_{T_e}$ (top panel) and $\mu_{0.66}$ vs. 12+$\log$(O/H)$_{T_e}$ (bottom panel) diagrams with our galaxies and the local $M_{\star}$--SFR stacks of \citet{Andrews2013}. 
The red squares denote our $z\sim 2$ galaxies, while the red filled squares are our $z\sim 2$ galaxies in the reliable sample.
The numbers written on the red squares are the ID numbers of our galaxies (Table \ref{tab:balmer_lines}).
The star mark represents the S16-stack.
The circles indicate the local $M_{\star}$--SFR stacks.
The triangles denote the GPs \citep{Amorin2012, Jaskot2013} whose $M_{\star}$ values are taken 
from \citet{Izotov2011} and \citet{Overzier2009}.
All of the symbols, the squares, star, circles, and triangles are color-coded by SFRs defined with the color bars at the right sides of the panels. 
The $z\sim2$ galaxies with EW$_0$({\Lya}) $>20$ {\AA} are marked with the large black squares.
In the bottom panel, the black dotted line represents the linear fit to the local $M_{\star}$--SFR stacks 
\citep{Andrews2013}. 
The data point of {\CSWA} is not shown because a $M_{\star}$ measurement of {\CSWA} is not available.
\label{fig:mz}}
\end{figure}

The top panel of Figure \ref{fig:mz} shows the mass--metallicity ($M_{\star}$--O/H) relation for our $z\sim2$ galaxies and the local $M_{\star}$--SFR stacks.
For comparison, we also plot nine green pea galaxies (GPs) with the triangles. These nine GPs are composed of three normal GPs \citep{Amorin2012} and six extreme GPs with high O$_{32}$ values \citep{Jaskot2013}.
We find that most of our $z\sim2$ galaxies, the S16-stack, and the nine GPs have 12+log(O/H) values comparable with those of the local stacks for given $M_{\star}$ and SFR values. 

The bottom panel of Figure \ref{fig:mz} compares our $z\sim2$ galaxies, the S16-stack, the local stacks, and GPs in the mass--metallicity--SFR ($M_{\star}$--O/H--SFR) relation.
Here the parameter $\mu_{0.66}$ is defined by
\begin{equation}
	\mu_{0.66} = \log(M_{\star}) - 0.66\log({\rm SFR}),
\end{equation}
which is introduced by \citet{Mannucci2010}. The coefficient of 0.66 is obtained by \citet{Andrews2013} to minimize the scatter of the local $M_{\star}$--SFR stacks. 
Seven out of the eleven $z\sim2$ galaxies, the S16-stack, and the nine GPs have 12+log(O/H) values comparable with those of the local stacks at fixed $\mu_{0.66}$.
The $\mu_{0.66}$ values of {\Lynx}, {\MACS}, and {\AbellS} range in $5\lesssim \mu_{0.66}\lesssim7$, where we find no counterparts in the local $M_{\star}$--SFR stacks. Thus, we cannot compare these three galaxies with the local $M_{\star}$--SFR stacks. It should be noted that these three galaxies are located above the extrapolation of the best fit line of the local stacks.

\subsection{N/O Abundance Ratio} \label{subsec:evolutionNO} 

\subsubsection{Overall Comparisons} \label{subsec:overallNO} 

\begin{figure*}
\begin{center}
\includegraphics[scale=0.6]{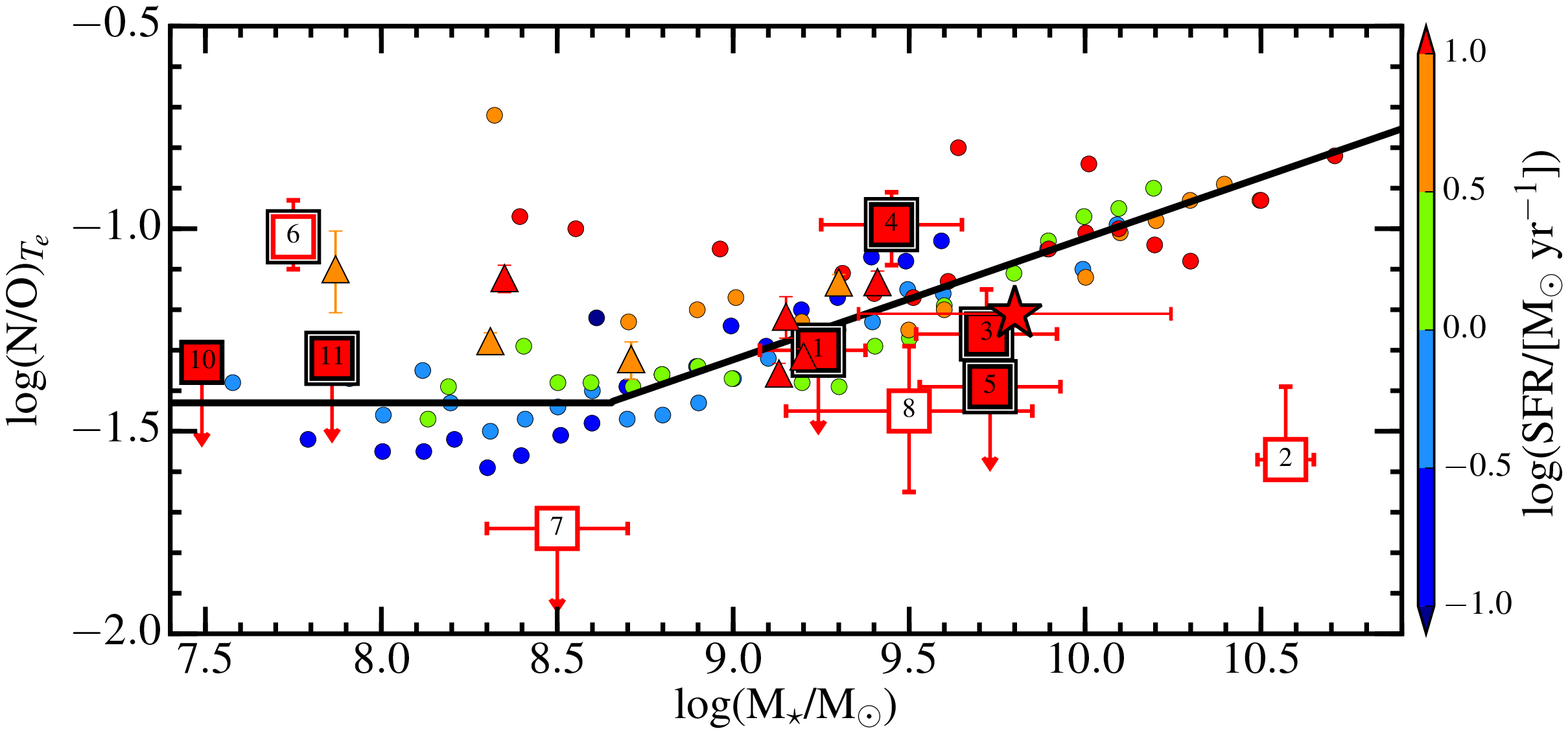}
\includegraphics[scale=0.6]{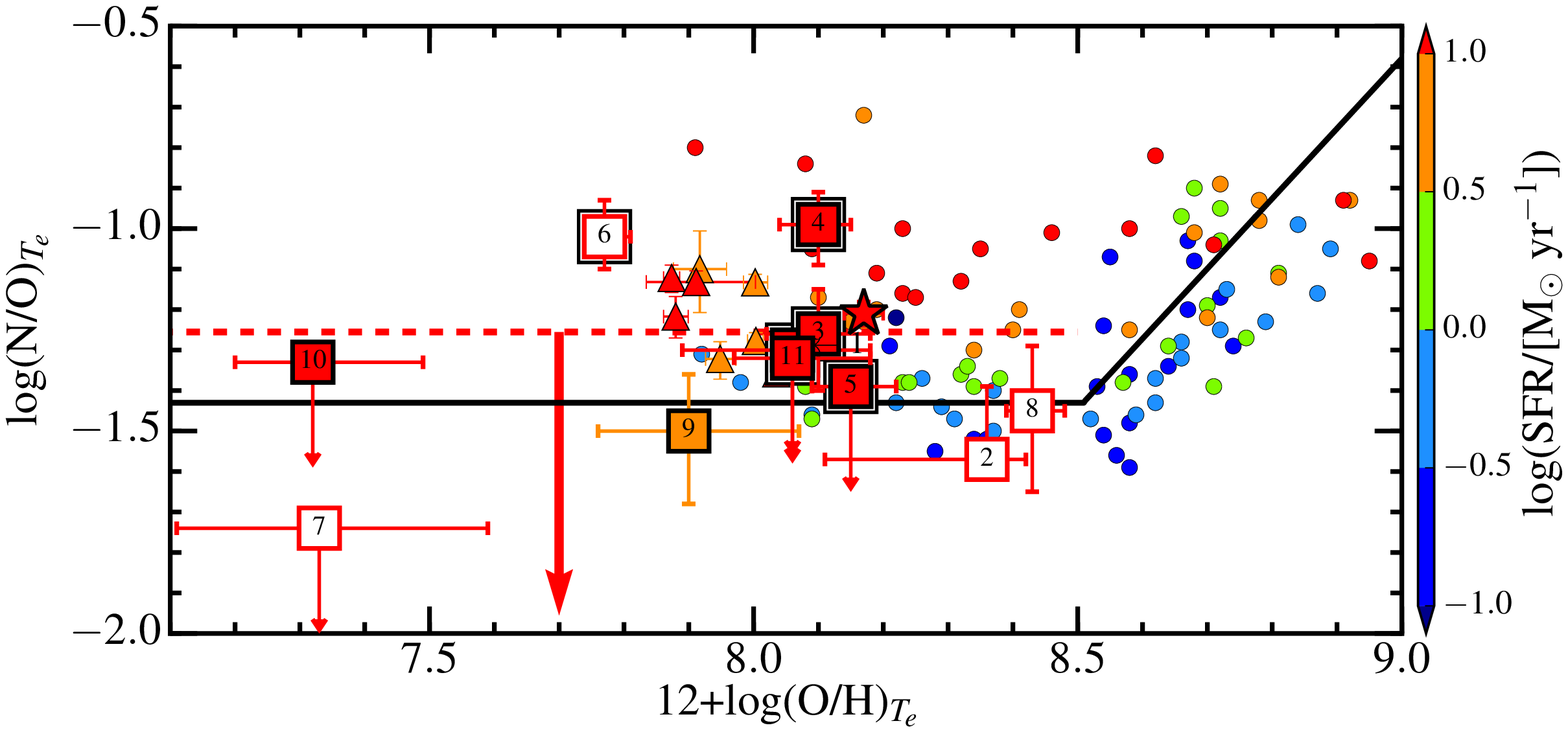}
\end{center}
\caption{Diagrams of $\log$(N/O)$_{T_e}$ vs. $\log(M_\star/M_\odot)$ (top panel) and $\log$(N/O)$_{T_e}$ vs. $12+\log$(O/H)$_{T_e}$ (bottom panel).
The squares represent our $z\sim 2$ galaxies, while the filled squares are our $z\sim 2$ galaxies in the reliable sample.
The star mark indicates the S16-stack.
The circles show the local $M_{\star}$--SFR stacks of \citet{Andrews2013}.
The triangles denote the GPs \citep{Amorin2012, Jaskot2013}.
All of the symbols, squares, star, circles, and triangles, have the color coding of SFR whose definition is the same as Figure \ref{fig:mz}.
The $z\sim2$ galaxies with EW$_0$({\Lya}) $>20$ {\AA} are marked with the large black squares.
The black solid lines represent the local sequence of \citet{Andrews2013}.
Although the black lines do not appear at the center of the distribution of the circles, 
the black lines represent the average values of the local stacks that are weighted by the numbers of galaxies (see text).
In the bottom panel, the red dashed line represents the upper limit of the average N/O value given by our $z\sim 2$ galaxies ($\log$(N/O)$_{T_e}=-1.26$). 
In the top panel, the data point of {\CSWA} is not shown because $M_{\star}$ of {\CSWA} is not available.  
All of the upper limits of $\log$(N/O)$_{T_e}$ indicated with the arrows correspond to the $2\sigma$ levels. 
\label{fig:no}}
\end{figure*}

Figure \ref{fig:no} presents $\log$(N/O)$_{T_e}$ values of $z\sim2$ galaxies as functions of $\log(M_\star/M_\odot)$ and 12+log(O/H)$_{T_e}$ in the top and bottom panels, respectively. 
We find that our galaxies range in $\log$(N/O)$_{T_e}\lesssim-1.0$, 12+log(O/H)=$7.3$--$8.5$, and $\log(M_\star/M_\odot)=7.5$--$10.5$.
We also show the S16-stack (star mark), the local stacks (circles), and the nine GPs (triangles).
The best-fit results of the local $M_{\star}$ stacks are indicated with the black lines in Figure \ref{fig:no}. 
Although the black lines do not appear at the center of the distribution of the circles in Figure \ref{fig:no}, 
the black lines represent the average of the local stacks that is weighted with the numbers of galaxies making the local stacks. 
We refer to these black lines as a ``local sequence''. 
Although there are many N/O--O/H relations obtained in various objects \citep[e.g.,][]{Groves2004, Perez-Montero2009, Pilyugin2012a}, we only exhibit the local stacks made by \citet{Andrews2013} for a comparison.
This is because our $z\sim2$ galaxies should be compared with typical $z\sim0$ SFGs whose N/O values are also based on the direct $T_e$ method. 
Because the local stacks are selected homogeneously, the local stacks are representative of typical $z\sim0$ SFGs.
Moreover, the physical properties of the local stacks are determined by the direct $T_e$ method.
Below we make two comparisons with the local sequence.

First, we compare our $z\sim2$ galaxies with the local sequence on an {\it individual} basis. 
In the top panel of Figure \ref{fig:no}, we present the N/O-$M_{\star}$ plot for $z\sim2$ galaxies whose N/O values are derived by the direct $T_e$ method. Most of our $z\sim2$ galaxies have $\log$(N/O)$_{T_e}$ values comparable with or lower than those of the local sequence.
Some objects have N/O values smaller than the local sequence beyond the errors.
This trend supports the previous observational results for $z\sim 2$ galaxies based on the strong line method \citep{Kashino2016,Strom2016}. \citet{Masters2016} suggest that N/O decrements in the N/O--$M_{\star}$ relation can be explained by the $z\sim2$ stellar population younger than the $z\sim0$ stellar population for a given $M_{\star}$.
In the bottom panel of Figure \ref{fig:no}, {\BXI}, {\BXII}, {\Lynx}, the S16-stack, and most of the GPs have $\log$(N/O)$_{T_e}$ values about $0.2$--$0.4$ dex larger than that of the local sequence ($\log$(N/O)$_{T_e}=-1.43$) significantly beyond the measurement errors. 
These three galaxies and the S16-stack are N/O-excess galaxies, where the word of N/O excess is hereafter used for objects that fall above the local sequence of the N/O--O/H relation.
The other eight $z\sim2$ galaxies have N/O values comparable to the local sequence. 

Second, we compare the {\it average} of our $z\sim2$ galaxies with the local sequence in the N/O--O/H relation.
We calculate the average value of N/O, using all of the eleven $z\sim2$ galaxies and the S16-stack with the weights of galaxy numbers and measurement uncertainties.
Because 5 out of our eleven $z\sim2$ galaxies have upper limits on N/O, 
we only derive an upper limit of the N/O average.
We thus obtain the stringent upper limit on the average N/O value.
The obtained average upper limit of our eleven galaxies and the S16-stack is $\log$(N/O)$_{T_e}<-1.26$. 
We show the average upper limit with the red dashed line in the bottom panel of Figure \ref{fig:no}. 
The average upper limit is higher than the local sequence by 0.17 dex in N/O, suggesting that the N/O ratios do not increase from $z\sim 0$ to $z\sim 2$ by more than 0.17 dex on average
for a given O/H. 

From these comparisons above, we confirm that most of our galaxies have N/O values 
comparable with (or lower than) that of the local sequence in the N/O--$M_{\star}$ relation, although some of the $z\sim2$ galaxies show the significant N/O excess above the local sequence in the N/O--O/H relation.
In Section \ref{sec:intro}, we have discussed the two possible physical origins of the N/O excess: the pristine gas inflow and the nitrogen enrichment by the enhanced WR population. The pristine gas inflow is likely to cause O/H decrements while N/O and $M_{\star}$ values change very little \citep{Andrews2013}, which means that an N/O excess appears only in the N/O--O/H relation. If the enhanced WR population enriches ISM with nitrogen, N/O values are likely to increase both in the N/O--$M_{\star}$ and N/O--O/H relation at $z\sim2$. Thus, our results support the pristine-gas-inflow scenario rather than the WR-star scenario. 
In addition, we confirm above that an N/O average does not increase very largely toward $z\sim2$ in the N/O--O/H relations for our sample. This may be primarily because our $z\sim2$ galaxies consist of relatively low-$M_{\star}$ galaxies whose N/O values are intrinsically smaller than the massive ones. Additionally, N/O decrements in the N/O--$M_{\star}$ relation towards $z\sim2$ may also contribute to the low N/O value on average.

\subsubsection{Comparisons of the Galaxies in the Same Ranges of SFR and $M_{\star}$} \label{subsec:fairNO} 

\begin{figure*}
\begin{center}
\includegraphics[scale=0.6]{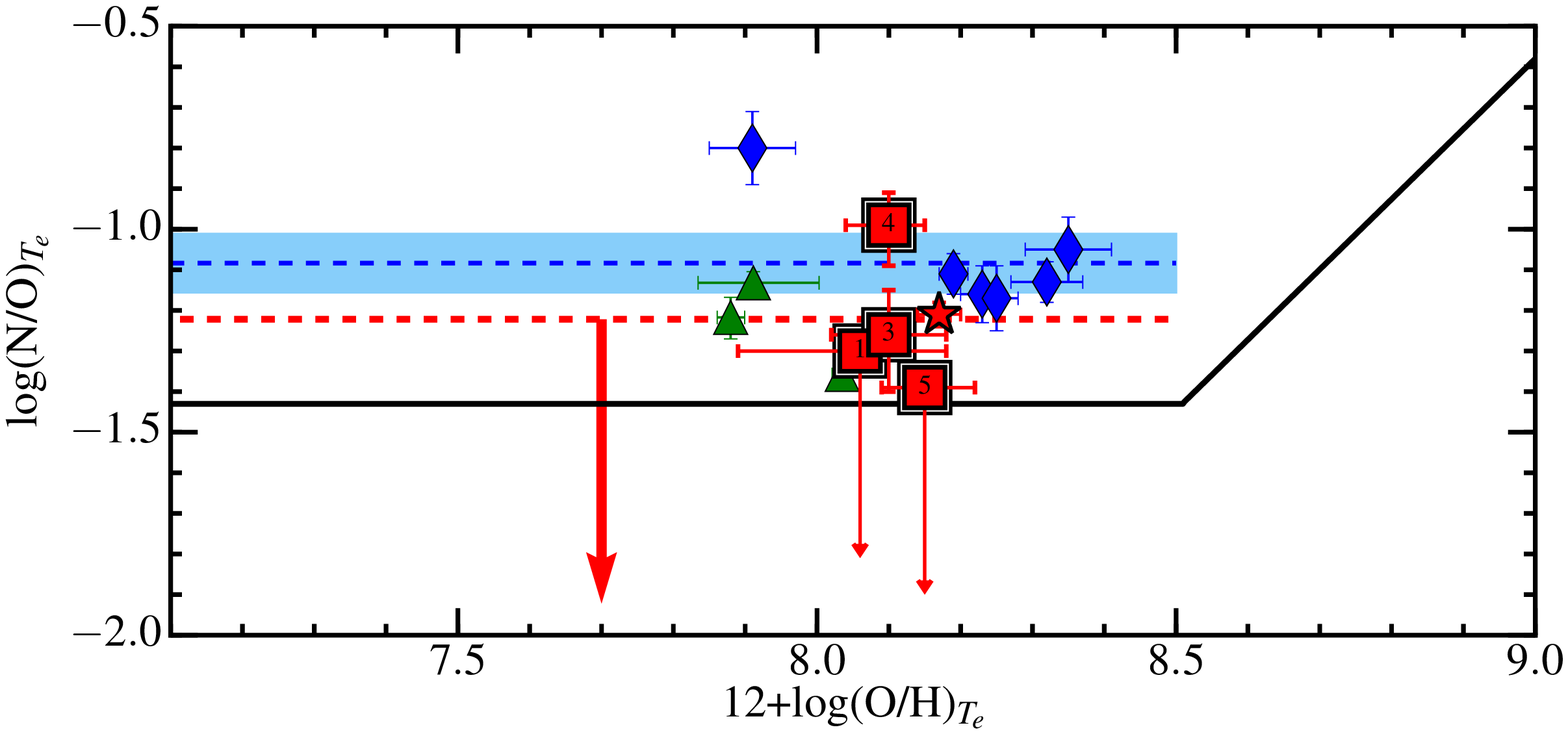}
\end{center}
\caption{Same as the bottom panel of Figure \ref{fig:no}, but for the galaxies in the parameter ranges of $\log$(SFR/M$_{\odot}$\,yr$^{-1}$)$>1.0$ and $9.0<\log(M_{\star}/M_{\odot})<10.0$. The blue diamonds and the green triangles indicate the LMHSs (Section \ref{subsec:fairNO}) and the GPs \citep{Amorin2012, Jaskot2013}, respectively. 
The red dashed line indicates the upper limit of averaged N/O value that is derived from {\COSMOS}, {\BXI}, {\BXII}, {\BXIII}, and the S-16 stack.
The blue dashed line denotes $\log$(N/O)$_{T_e}=-1.08$, which is the average of LMHSs with the weight of the number of galaxies,
and the blue shaded region represents the 1 $\sigma$ range ($\pm 0.07$ dex) for the determination of the blue dashed line.
\label{fig:no_oh_limit}}
\end{figure*}

Figure \ref{fig:no} shows that most of the local stacks with high SFRs (red circles) have large N/O ratios of 
$\log$(N/O)$_{T_e}$=$-1.2$ to $-0.8$. 
This is the dependence of N/O ratio on SFR that is claimed by \citet{Andrews2013}. Similarly, the top panel of Figure \ref{fig:no} exhibits a clear dependence of N/O ratio on $M_{\star}$ for the local stacks. Because N/O ratios of local galaxies correlate with SFR and $M_{\star}$, one needs to compare N/O ratios of local and $z\sim2$ galaxies in the same ranges of SFR and $M_{\star}$ values.

We find that there is an overlap of the $z\sim 2$ and local galaxies in the parameter ranges of 
$\log$(SFR/M$_{\odot}$\,yr$^{-1}$)$>1.0$ and $9.0<\log(M_{\star}/M_{\odot})<10.0$. 
We choose galaxies in these ranges from our $z\sim2$ galaxies and the local $M_{\star}$--SFR stacks. 
These are galaxies from 
our reliable sample ({\COSMOS}, {\BXI}, {\BXII}, and {\BXIII}), the S16-stack, and six local $M_{\star}$--SFR stacks. The six local $M_{\star}$--SFR stacks are made of 206 SDSS galaxies in total. 
We refer to these six local $M_{\star}$--SFR stacks as low-mass and high-SFR stacks (LMHSs).

Figure \ref{fig:no_oh_limit} presents $\log$(N/O)$_{T_e}$ as a function of 12+log(O/H)$_{T_e}$ for {\COSMOS}, {\BXI}, {\BXII}, {\BXIII}, and the six LMHSs. 
Figure \ref{fig:no_oh_limit} also shows four out of the nine GPs that fall in the ranges of $\log$(SFR/M$_{\odot}$\,yr$^{-1}$)$>1.0$ and $9.0<\log(M_{\star}/M_{\odot})<10.0$.
The $\log$(N/O)$_{T_e}$ values of {\BXI} and {\BXII} are significantly higher than the local sequence by $\sim 0.2$--$0.4$ dex, but {\COSMOS} and {\BXIII} have $\log$(N/O)$_{T_e}$ upper limits comparable with the local sequence. 
The $\log$(N/O)$_{T_e}$ average of 
our sample galaxies
is comparable with or lower than those of GPs and LMHSs.

\subsection{Ionization Parameter} \label{subsec:evolutionq} 

\begin{figure*}
\begin{center}
\includegraphics[scale=0.75]{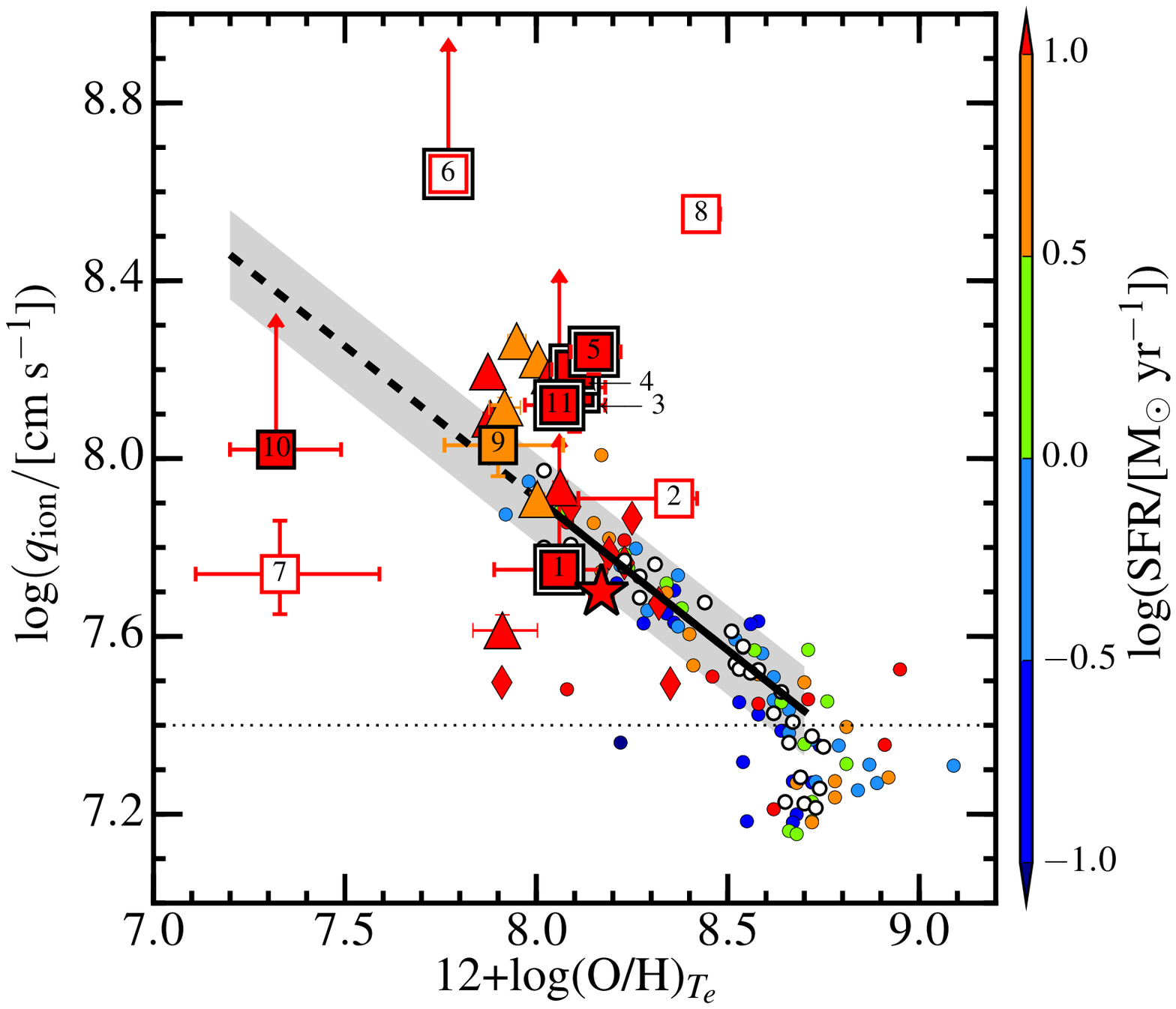}
\end{center}
\caption{$\log(q_{\rm ion})$ as a function of $12+\log$(O/H)$_{T_e}$.
The squares represent our $z\sim 2$ galaxies, while the filled squares show our $z\sim 2$ galaxies in the reliable sample.
The squares with the arrows indicate the lower limits of {\q} at the $2\sigma$ levels.
The star mark indicates the S16-stack.
The circles show the local $M_{\star}$--SFR stacks of \citet{Andrews2013}.
The triangles and diamonds denote the local galaxies of GPs \citep{Amorin2012, Jaskot2013} and LMHSs (Section \ref{subsec:fairNO}).
All of the symbols, squares, circles, triangles, and diamonds have the color coding of SFR
whose definition is the same as Figure \ref{fig:mz}. 
The $z\sim2$ galaxies with EW$_0$({\Lya}) $>20$ {\AA} are marked with the large black squares.
The black solid line (the gray shaded region) indicates the best-fit linear function (the uncertainty) for the local $M_{\star}$ stacks. 
The black dotted line represents $\log$(\q)$=7.4$, above which we perform the linear fitting. 
The black dashed line and the associated gray shade denote the extrapolation of the best-fit linear function and the $0.1$ dex scatter in {\q}. 
\label{fig:zq}}
\end{figure*}

Figure \ref{fig:zq} presents the {\q} values of 
our $z\sim2$ galaxies, the S16-stack, 
the local stacks, and the nine GPs as a function of 12+log(O/H). 
The local stacks indicate the tight anti-correlation between 12+log(O/H) and {\q} that are measured by the direct $T_e$ method. This anti-correlation has been reported by \citet{Nakajima2014} and \citet{Onodera2016}, who estimate the O/H and {\q} values based on the strong line methods.
The black line in Figure \ref{fig:zq} indicates the average of the local stacks that is the best-fit linear function for the local  $M_{\star}$ stacks (open circle) in the range of $\log(q_{\rm ion})>7.4$. The best-fit linear function is
\begin{equation}
	\log(q_{\rm ion})=-0.68(12+\log({\rm O/H})_{T_e})+13.38. \label{eq:zq}
\end{equation}
Six out of the nine GPs show {\q} values higher than that of the local-stack best-fit function, although GPs are parts of $z\sim0$ galaxies.

For our sample, seven out of our eleven $z\sim2$ galaxies ({\SMACS}, {\BXI}, {\BXII}, {\BXIII}, {\Lynx}, {\SGAS}, and {\AbellS}) have {\q} values falling above the local-stack best-fit function, 
which are hereafter refereed to as {\q} excesses.
These {\q} values of the seven $z\sim2$ galaxies are similar to those of the GPs with {\q} excesses. 
This result does not change if we compare our reliable-sample galaxies with the GPs.
Because we find no {\q} dependence on SFR in Figure \ref{fig:zq}, the high {\q} values may not be attributed to high SFRs. 
It should be noted that the {\q} excess can be mimicked by the harder spectrum of ionizing radiation (see Section \ref{sec:intro}) or the density-bounded {\HII} regions \citep{Nakajima2014}.

\begin{figure}
\begin{center}
\includegraphics[scale=0.50]{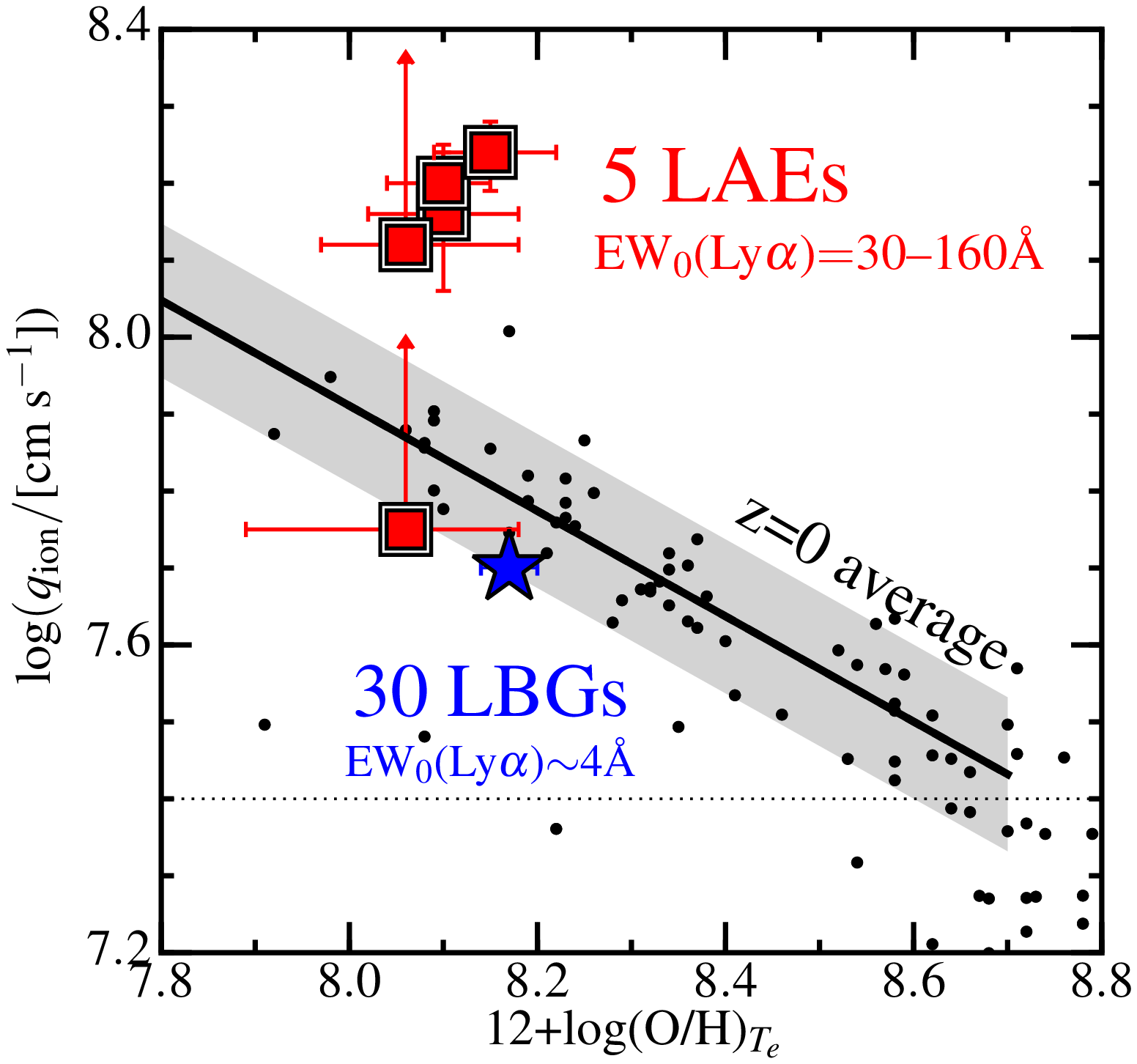}
\end{center}
\caption{Same as the bottom panel of Figure \ref{fig:zq}, but for the reliable galaxies with the measurements of EW$_0$({\Lya}).
The red and blue colors indicate galaxies with EW$_0$({\Lya})$>20${\AA} (classified as LAEs) and EW$_0$({\Lya})$<20${\AA}, respectively.
\label{fig:zq_limit}}
\end{figure}

In the reliable sample, {\COSMOS}, {\BXI}, {\BXII}, {\BXIII}, {\AbellS}, and the S16-stack have EW$_0$({\Lya}) estimates (Table \ref{tab:prop}), because {\Lya} lines are covered in their spectra.
The five galaxies of {\COSMOS}, {\BXI}, {\BXII}, {\BXIII}, and {\AbellS} have EW$_0$({\Lya}) values of $30$--$160${\AA}, which are classified as LAEs (defined by EW$_0$({\Lya})$>$20{\AA}).
In contrast, the S16-stack has a low EW$_0$({\Lya}) value of $\sim4${\AA}.
This low EW$_0$({\Lya}) value is explained by the fact that the S16-stack is composed of UV-continuum bright galaxies.
In Figure \ref{fig:zq_limit}, we find that the majority of the five LAEs show significant {\q} excesses, while the S16-stack has no {\q} excess.
This result suggests the possibility that there is a positive correlation between {\q} and EW$_0$({\Lya}),
and support the claim of \citet{Nakajima2014}. Recently, \citet{Trainor2016} also report that Lyman break galaxies (LBGs) with high EW$_0$({\Lya}) values show high O$_{32}$ values,
indicative of the positive correlation between {\q} and EW$_0$({\Lya}).

\subsection{Physical Origins of the BPT Diagram Offset} \label{subsec:BPT} 

\begin{figure*}[]
\begin{center}
\includegraphics[scale=0.65]{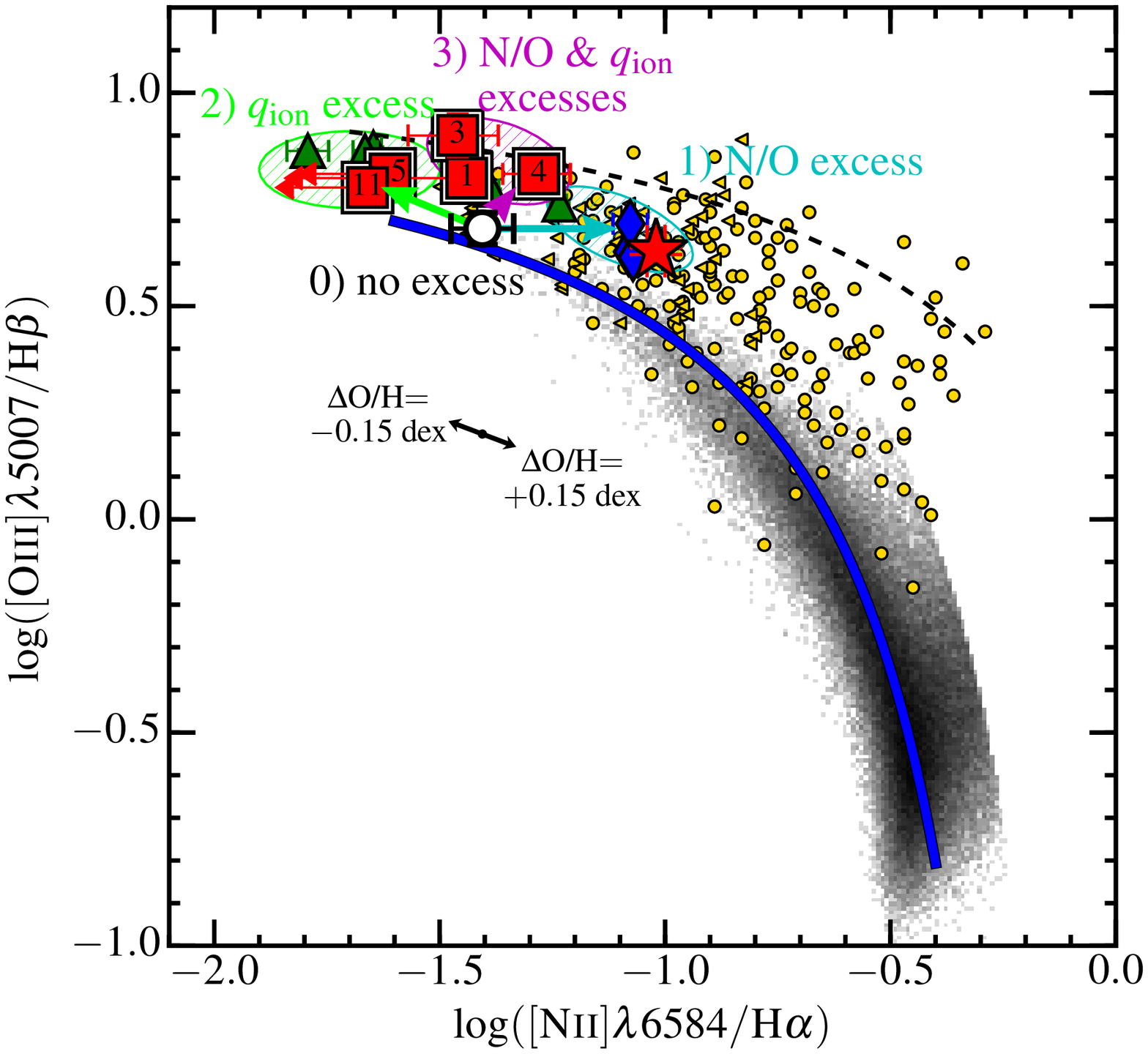}
\end{center}
\caption{%
Same as Figure \ref{fig:bpt}, but for the five galaxies (red squares) and the S16-stack (red star mark).
We compare the GPs (green triangles) and LHMSs (blue diamonds).
All of these galaxies and the stack have similar O/H values falling in 12+log(O/H)$=8.10\pm0.15$. 
The open circle represents the reference galaxy with 12+log(O/H)$=8.10$ that shows neither N/O nor {\q} excess (see text). 
The $z\sim2$ galaxies with EW$_0$({\Lya}) $>20$ {\AA} are marked with the large black squares.
The black arrows indicate the changes of the reference galaxy position when the oxygen abundances are increased/decreased by $0.15$ dex which are calculated by the \textsc{Cloudy} models.
Here we also present the positions of the three groups of 1), 2), and 3) with the cyan, green, and magenta
shades, respectively.
The cyan, green, and magenta arrows indicate the differences of $\log$({\NII}/{\Ha}) and $\log$({\OIII}/{\Hb}) values 
given by our \textsc{Cloudy} modeling for galaxies with 0.3 dex excesses of 1) $\log$(N/O)$_{T_e}$, 2) {\q}, and 3) $\log$(N/O)$_{T_e}$ $+$ {\q}, respectively,
where we assume 12+log(O/H)$=8.10$.
\label{fig:bpt_comp}}
\end{figure*}

In this section, we discuss the positions of the galaxies on the BPT diagram.
Figure \ref{fig:bpt_comp} presents the BPT diagram,
which is the same as Figure \ref{fig:bpt}, but for 
five galaxies (ID-1, 3, 4, 5, and 11) selected from our reliable sample. 
Similarly, S16-stack is also presented.
These five galaxies and the S16-stack have very similar oxygen abundances,
which fall in 12+log(O/H)$=8.10\pm0.15$.
Two out of the five galaxies ({\BXI} and {\BXII}) and the S16-stack significantly exhibit the BPT offsets,
while the other three galaxies, {\COSMOS}, {\BXIII} and {\AbellS}, have signatures of possible BPT offsets with {\NII}/{\Ha} upper limits.
We also plot the five GPs and the three LMHSs 
\footnote{
The three LMHSs are 24, 18, and 4 individual galaxy stacks,
and composed of a total of 46 (=24+18+4) individual galaxies.
}
that have O/H values of 12+log(O/H)$=8.10\pm0.15$. 
In Figure \ref{fig:bpt_comp}, we present the reference galaxy (open circle),
which is an average local galaxy with 12+log(O/H)$=8.10$ that shows neither N/O nor {\q} excess.
More precisely, the reference galaxy is the average of four local stacks (a total of 46 local galaxies) 
with a metallicity of 12+log(O/H)$=8.10\pm0.05$ and no N/O+{\q} excesses.
The reference galaxy has N2 and O3 values comparable with the best-fit sequence of the SDSS galaxies \citep{Kewley2008} within the error bar.

In Figure \ref{fig:bpt_comp}, we find that the $z\sim2$ galaxies and the GPs are located very close on the BPT diagram, which is consistent with previous studies  \citep[e.g.][]{Steidel2014}. Because the locations on the BPT diagram are degenerated with N/O-excess galaxies and {\q}-excess galaxies at different O/H values, no previous study has clearly identified the physical origin of the BPT offsets of $z\sim2$ galaxies. Our simultaneous measurements of N/O, O/H, and {\q} values without systematic uncertainties can break the degeneracy between O/H, N/O, and {\q} excesses (Sections \ref{subsec:evolutionNO} and \ref{subsec:evolutionq}).

From the results in Sections \ref{subsec:evolutionNO} and \ref{subsec:evolutionq}, the galaxies with 12+log(O/H)$=8.10\pm0.15$ are classified into four groups based on the N/O and/or {\q} excesses as shown below:
\begin{enumerate}
\item[0)] Galaxies with {\it neither} N/O nor {\q} excess.
\item[1)] Galaxies with an N/O excess and {\it no} {\q} excess.
\item[2)] Galaxies with a {\q} excess and {\it no} N/O excess.
\item[3)] Galaxies with both N/O and {\q} excesses.
\end{enumerate}
The classifications of the $z\sim2$ galaxies, S16-stack, GPs, and LMHSs are summarized in Table \ref{tab:class}. 

\input tab6.tex

In Figure \ref{fig:bpt_comp}, the galaxies classified into 1), 2), and 3) are surrounded by the cyan, light green, and magenta circles on the BPT diagram, respectively. We find these three groups of galaxies are located separately on the BPT diagram regardless of the redshift. 
We argue that the BPT offsets  are caused by the N/O and {\q} excesses while the amounts of the contributions of N/O and {\q} are different from galaxy to galaxy.

We construct \textsc{Cloudy} photoionization models to confirm the relation between the physical properties and the loci on the BPT diagram.
Assuming 12+log(O/H)$=8.10$, we calculate the differences of {\NII}/{\Ha} and {\OIII}/{\Hb} values with the N/O and {\q} excesses by $0.3$ dex
from the reference galaxy.
These N/O and {\q} excesses by $0.3$ dex are roughly similar to those of the galaxies in 1), 2), and 3). 
In Figure \ref{fig:bpt_comp}, the cyan, light green, and magenta arrows indicate the {\NII}/{\Ha} and {\OIII}/{\Hb} differences 
that are estimated by our \textsc{Cloudy} modeling. The cyan (light green) arrow corresponds to the N/O ({\q}) excess by 0.3 dex 
and the magenta arrow indicates both N/O and {\q} excesses by 0.3 dex.
We confirm that these cyan, light green, and magenta arrows point to the positions of the 1), 2), and 3) galaxy groups, respectively.
Thus, the positions of the 1), 2), and 3) galaxies can be explained by the excesses in N/O, {\q}, and N/O $+$ {\q}, respectively. 

In these arguments, we assume that all of the galaxies, our galaxies, S16-stack, GPs, and LMHSs, 
have exactly the same oxygen abundance of 12+log(O/H)$=8.10$. However, the oxygen abundances of these galaxies 
fall in the finite range of 12+log(O/H)$=8.10\pm0.15$.
We examine how much the reference galaxy changes the position in Figure \ref{fig:bpt_comp}
for the oxygen abundances from 12+log(O/H)$=8.10-0.15$ to $8.10+0.15$.
We make two \textsc{Cloudy} models that are the same as 
the reference galaxy, but with the oxygen abundances of 12+log(O/H)=$8.10-0.15$ and $8.10+0.15$.  
We find that these two \textsc{Cloudy} models give {\NII}/{\Ha} and {\OIII}/{\Hb} values that differ from the 
reference galaxy only by $\sim 0.1$ dex and $\sim 0.03$ dex, respectively.
In Figure 11, we show these differences of the two models 
from the reference model with the black arrows. 
The oxygen abundance differences by $\pm 0.15$ dex gives negligible changes in
Figure \ref{fig:bpt_comp}.

We find that a large fraction of our $z\sim 2$ galaxies ({\COSMOS}, {\BXIII}, {\AbellS}, {\BXI}, {\BXII} , and the S16-stack) 
with BPT offsets have {\q} and/or N/O excesses, which are revealed by
the direct $T_e$ method, for the first time.
Moreover, $z\sim 2$ galaxies with BPT offsets are not drawn from one of 1)--3) galaxy populations alone,
but composed of 1)--3) populations.

We find that the 1)--3) population galaxies also exist at $z\sim 0$. However, these population galaxies are very rare in the local galaxies. 
In fact, we find that only less than 2,000 local galaxies indicate N/O or {\q} excesses in the 208,529 SDSS galaxies ($<$1\%).
 These results suggest that 1)--3) population galaxies with BPT offsets emerge at $z\sim 2$ 
 and mostly disappear in the present-day universe, except for local galaxies such as GPs and LMHSs.

\section{SUMMARY} \label{sec:summary}
We have investigated the oxygen abundances (O/H), nitrogen-to-oxygen abundance ratios (N/O), and the ionization parameters ({\q}) for the eleven $z\sim2$ galaxies and a stack made of 30 KBSS galaxies (dubbed ``S16-stack'') with {\OIII} $\lambda$4363 and {\OIIIsemi} $\lambda \lambda$1661,1666 lines. 
Our $z\sim2$ galaxy sample includes {\Lya} emitters (LAEs) and Lyman break galaxies (LBGs).
We have measured the O/H, N/O, and {\q} values by the direct $T_e$ method, and studied the evolution of N/O and {\q} from $z\sim 0$ to $z\sim 2$.
We have compared our $z\sim2$ galaxies with the local stacks of the 208,529 SDSS galaxies and 9 GP galaxies, whose ISM properties are also measured by the direct $T_e$ method.
Our main results are summarized below. 

\begin{itemize}

\item We have shown, for the first time, the relations of N/O-$M_{\star}$ and N/O-O/H for the $z=0$--$2$ galaxies with the measurements by the direct $T_e$ method. 
In the N/O-$M_{\star}$ relation (the top panel of Figure \ref{fig:no}), 
the majority of the $z\sim 2$ galaxies have the N/O ratios comparable with (or lower than) the local sequence. 
In the N/O-O/H relation (the bottom panel of Figure \ref{fig:no}), 
we find that there exist $z\sim 2$ galaxies with an N/O excess falling beyond the local sequence. 
For given O/H, our sample places the upper limit of N/O ratio $\log ({\rm N/O})\le -1.26$ on average
for $z\sim 2$ galaxies, suggesting that the N/O ratio does not increase largely ($<0.17$ dex). 

\item We have compared the N/O values of the $z\sim2$ and $z\sim0$ galaxies in the same SFR and $M_{\star}$ ranges ($\log$(SFR/$M_{\odot}$ yr$^{-1}$)$>1.0$ and $\log(M_{\star}/M_{\odot})=9.0$--$10.0$). We choose our four $z\sim2$ galaxies, the S16-stack, 206 SDSS galaxies (dubbed LMHSs), and four GPs. The N/O average of $z\sim2$ galaxies and the S16-stack are comparable to or lower than those of GPs and LMHSs (Figure \ref{fig:no_oh_limit}).

\item We have presented, for the first time, the relation {\q}-O/H for the $z=0$--$2$ galaxies with the measurements by the direct $T_e$ method. We have identified that the seven out of our eleven galaxies at $z\sim2$ have {\q} excesses at fixed O/H (Figure \ref{fig:zq}). 
These {\q} excesses are similar to those of some of the GPs in the local universe (Figure \ref{fig:zq}). 
In our reliable sample, the majority of $z\sim2$ galaxies with EW$_0$({\Lya})$>20${\AA} indicate significant {\q} excesses, while the S16-stack (EW$_0$({\Lya})$\sim4${\AA}) shows no {\q} excess (Figure \ref{fig:zq_limit}). Thus we claim that {\q} excesses are related to high EW$_0$({\Lya}) values.

\item We have selected galaxies within 12+log(O/H)$=8.10\pm0.15$ and compared their loci on the BPT diagram (Figure \ref{fig:bpt_comp}). By fixing O/H values, we have resolved the degeneracies between N/O and {\q} on the BPT diagram. 
These $z\sim 2$ galaxies show BPT offsets, and the direct $T_e$ method measurements
indicate that these are 
1) galaxies with an N/O excess and {\it no} {\q} excess, 
2) galaxies with a {\q} excess and {\it no} N/O excess, and 
3) galaxies with N/O and {\q} excesses,
where the words of N/O and {\q} excesses are defined by the positive departures of N/O and {\q} measurements, respectively, from the local sequences for a fixed O/H. 
These results suggest that the BPT offsets at $z\sim 2$ are not made by 
one of 1)--3) galaxy populations alone, but the composite of 1)--3) populations.
We identify that local galaxies of LMHSs and GPs with BPT offsets similarly have 1)--3) properties.
These local galaxies are probably counterparts of typical $z\sim 2$ galaxies.
  
\item A large fraction of our $z\sim 2$ galaxies are classified as 1)--3) galaxies with N/O and/or {\q} excesses, while such N/O- and/or {\q}-excess galaxies are very rare in the local galaxies ($<$ 1\%). These results would indicate that galaxies with the BPT offsets emerge at $z\sim2$ and mostly disappear at $z\sim0$, except for local galaxies such as GPs and LMHSs.
\end{itemize}

\bigskip
We thank the anonymous referee for constructive comments and suggestions.
We are grateful to 
Mark Dijikstra,
Dawn Erb,
Andrea Ferrara,
Tomotsugu Goto,
Daichi Kashino, 
Lisa Kewley,
Adam Leroy, 
Crystal Martin,
Daniel Masters,
Kentaro Motohara,
Tohru Nagao,
Masato Onodera,
Ryan Sanders,
Alice Shapley,
Rhythm Shimakawa,
Kazuhiro Shimasaku,
Charles Steidel,
Claudia Lagos Urbina,
and 
Kiyoto Yabe,
for their useful comments and discussions.

Funding for the SDSS and SDSS-II has been provided by the Alfred P. Sloan Foundation, the Participating Institutions, the National Science Foundation, the U.S. Department of Energy, the National Aeronautics and Space Administration, the Japanese Monbukagakusho, the Max Planck Society, and the Higher Education Funding Council for England. The SDSS Web Site is http://www.sdss.org/.

The SDSS is managed by the Astrophysical Research Consortium for the Participating Institutions. The Participating Institutions are the American Museum of Natural History, Astrophysical Institute Potsdam, University of Basel, University of Cambridge, Case Western Reserve University, University of Chicago, Drexel University, Fermilab, the Institute for Advanced Study, the Japan Participation Group, Johns Hopkins University, the Joint Institute for Nuclear Astrophysics, the Kavli Institute for Particle Astrophysics and Cosmology, the Korean Scientist Group, the Chinese Academy of Sciences (LAMOST), Los Alamos National Laboratory, the Max-Planck-Institute for Astronomy (MPIA), the Max-Planck-Institute for Astrophysics (MPA), New Mexico State University, Ohio State University, University of Pittsburgh, University of Portsmouth, Princeton University, the United States Naval Observatory, and the University of Washington.

This work is supported by World Premier International Research Center Initiative (WPI Initiative), MEXT,
Japan, and KAKENHI (15H02064) Grant-in-Aid for Scientific Research (A) through Japan 
Society for the Promotion of Science (JSPS).

\bibliography{library}

\end{document}

%% file: tab1.tex
\begin{table}[!hbtp]
\caption{Emission Line Fluxes of {\COSMOSno}}
\begin{center}
{\scriptsize
\begin{tabular}{cccc}
\hline\noalign{\vskip3pt} 
Line &
F$({\lambda})$\footnotemark[a] &
I$({\lambda})$\footnotemark[a] &
Instrument  \\ 
\noalign{\vskip3pt} 
(1) &
(2) &
(3) & 
(4) \\
\hline\noalign{\vskip3pt} 
{\OIIIsemi} $\lambda$1661& 
$1.48\pm0.37$ &
$1.70\pm0.43$ &
Keck/LRIS\\
{\OIIIsemi} $\lambda$1666& 
$2.29\pm0.45$ &
$2.63\pm0.52$ &
Keck/LRIS\\
{\NIIIsemi} $\lambda$1750& 
$<1.28$\footnotemark[b] &
$<1.47$\footnotemark[b] &
Keck/LRIS\\
{\OII} $\lambda$3727& 
$<88.3$\footnotemark[b] &
$<98.0$\footnotemark[b] &
Subaru/FMOS\\
{\Hb} $\lambda$4861& 
$30.5\pm9.9$ &
$33.2\pm10.8$ &
Subaru/FMOS\\
{\OIII} $\lambda$4959& 
$57.6\pm12.9$ &
$62.8\pm14.1$ &
Subaru/FMOS\\
{\OIII} $\lambda$5007& 
$197\pm8.7$ &
$215\pm9.5$ &
Subaru/FMOS\\
{\Ha} $\lambda$6563& 
$86.6\pm1.4$&
$91.8\pm1.5$&
Keck/MOSFIRE\\
{\NII} $\lambda$6584& 
$<3.06$\footnotemark[b]&
$<3.24$\footnotemark[b]&
Keck/MOSFIRE\\
\hline\noalign{\vskip3pt} 
\end{tabular}}
\end{center}
{\footnotesize 
\hangindent6pt\noindent
(1): Emission lines and the rest-frame wavelength. (2): Measured fluxes. (3): Dust-corrected fluxes under the assumption of $E(B-V)_{\rm neb}=0.02$ mag that are given by the Balmer decrement measurement (Section \ref{sec:sample}). 
(4): Instruments used in our observations.\\
\hbox to6pt{\footnotemark[a]\hss} In units of $10^{-18}$erg\,cm$^{-2}$\,s$^{-1}$.\\
\hbox to6pt{\footnotemark[b]\hss} $2\sigma$ upper limit.
}\label{tbl:c12805}
\end{table}

%% file: tab2.tex
\begin{landscape}
\begin{table}[!hbtp]
\caption{Physical Properties of the $\lowercase{z}\sim2$ Galaxies}
\begin{center}
{\scriptsize
\begin{tabular}{p{3pt}lcccccccccc}
\hline\noalign{\vskip3pt} 
ID &
Object &
$z$ &
$\log M_*$ &
SFR &
EW$_0$({\Lya}) &
$T_e$(O\textsc{iii}) &
12+$\log$(O/H)$_{T_e}$ &
$\log$(N/O)$_{T_e}$ &
$\log (q_{\rm ion})$ &
Reliable &
Ref.\\
&
&
&
($M_{\odot}$)&
($M_{\odot}$ yr$^{-1}$)&
(\AA)&
(10$^4$ K)&
&
&
(cm s$^{-1}$)&
&
 \\
(1)&
(2)&
(3)&
(4)&
(5)&
(6)&
(7)&
(8)&
(9)&
(10)&
(11)&
(12)\\
\hline\noalign{\vskip3pt} 
1&
\COSMOSno  & 
2.159 & 
9.24$^{+0.13}_{-0.17}$ & 
18 & 
33.7$\pm$6.0 &
$1.29^{+0.78}_{-0.09}$&
$8.06^{+0.12}_{-0.17}$&
$<-1.30$\footnotemark[a]\footnotemark[\dag]&
$>7.75$\footnotemark[a]&
{\it yes}&
\ldots \\
2&
\SMACSno  & 
1.963 & 
10.57$\pm$ 0.08 & 
16 & 
$\lesssim 0$ &
$1.03^{+0.03}_{-0.04}$&
$8.36^{+0.06}_{-0.25}$&
$-1.57^{+0.18}_{-0.03}$&
$7.91\pm0.03$&
{\it no}&
1\\
3&
\BXIno  & 
2.189 & 
9.72\footnotemark[b] & 
58.1\footnotemark[b] & 
101.9$\pm$4.4\footnotemark[b] &
$1.36^{+0.11}_{-0.09}$&
$8.10\pm0.08$&
$-1.26^{+0.11}_{-0.14}$&
$8.16^{+0.09}_{-0.10}$&
{\it yes}&
2, 3 \\
4&
\BXIIno  & 
2.305 & 
9.45\footnotemark[b] & 
52.0\footnotemark[b]& 
61.8$\pm$2.6\footnotemark[b] &
$1.28^{+0.06}_{-0.05}$&
$8.10^{+0.05}_{-0.06}$&
$-0.99^{+0.08}_{-0.10}$&
$8.20^{+0.04}_{-0.05}$&
{\it yes}&
2, 3 \\
5&
\BXIIIno  & 
2.174 & 
9.73\footnotemark[b] & 
28.8\footnotemark[b]& 
40.4$\pm$2.2\footnotemark[b] &
$1.27\pm0.07$&
$8.15^{+0.07}_{-0.06}$&
$<-1.39$\footnotemark[a] &
$8.24^{+0.04}_{-0.05}$&
{\it yes}&
2, 3\\
6&
\Lynxno  & 
3.357 & 
7.75\footnotemark[c]& 
97 & 
$\sim$480\footnotemark[d] &
$1.74^{+0.05}_{-0.04}$&
$7.77^{+0.04}_{-0.03}$&
$-1.02^{+0.09}_{-0.08}$\footnotemark[\dag]&
$>8.64$\footnotemark[a] &
{\it no}&
4, 5 \\
7&
\AbellYKno  & 
1.703 & 
8.5$\pm$0.2 & 
76 & 
\ldots &
$3.01^{+1.18}_{-0.83}$&
$7.33^{+0.26}_{-0.22}$&
$<-1.74$\footnotemark[a]&
$7.74^{+0.12}_{-0.09}$&
{\it no}&
6\\
8&
\SGASno  & 
3.625 & 
9.5$\pm$0.35 & 
84 & 
$\lesssim 0$ 
&
$1.05^{+0.03}_{-0.04}$&
$8.43^{+0.05}_{-0.04}$&
$-1.45^{+0.16}_{-0.20}$\footnotemark[\dag]&
$8.55^{+0.04}_{-0.03}$&
{\it no}&
7\\
9&
\CSWAno  & 
1.433 & 
\ldots & 
5.7 & 
\ldots &
$1.41^{+0.22}_{-0.19}$&
$7.90^{+0.17}_{-0.14}$&
$-1.50^{+0.14}_{-0.18}$&
$8.03^{+0.09}_{-0.07}$&
{\it yes}&
8, 9\\
10&
\MACSno & 
2.060 & 
7.49 & 
906 & 
\ldots &
$2.19^{+0.34}_{-0.36}$&
$7.32^{+0.17}_{-0.12}$&
$<-1.33$\footnotemark[a]\footnotemark[\dag]&
$>8.02$\footnotemark[a]&
{\it yes}&
10\\
11&
\AbellSno  & 
1.702 & 
7.86 & 
55 & 
163.8$\pm$25.5&
$1.28^{+0.10}_{-0.12}$&
$8.06^{+0.12}_{-0.09}$&
$<-1.32$\footnotemark[a]\footnotemark[\dag]&
$>8.12$\footnotemark[a]&
{\it yes} &
10\\
12&
\KBSS  & 
$2.396\pm0.111$\footnotemark[e]& 
$9.8\pm0.3$\footnotemark[e]& 
$29.2\pm17.6$\footnotemark[e]& 
$\sim4$\footnotemark[f]&
$1.21^{+0.03}_{-0.04}$&
$8.17\pm0.03$&
$-1.21\pm0.03$&
$7.70\pm0.02$&
{\it yes} &
11\\
\hline\noalign{\vskip3pt} 
\end{tabular}}
\end{center}
{\footnotesize 
(1): ID. (2): Name of the object. (3): Systemic redshift. (4): Stellar mass. The stellar masses are estimated by SED fitting, except for the {\Lynx}. (5): Star formation rate inferred from the Balmer line and the relation of \citet{Kennicutt1998}. (6): Rest-frame equivalent width of a {\Lya} emission line.  (7): Electron temperature of O$^{2+}$. (8): Oxygen abundance derived from the direct $T_e$ methods. (9): Nitrogen to oxygen abundance ratio calculated by the direct $T_e$ methods.  (10): Ionization parameter estimated with the \OIII/\OII\, line ratio \citep{Kewley2002}. (11): {\it yes} ({\it no}) indicates that the object is (is not) a galaxy in the reliable sample. (12): Reference. 1: \citet{Christensen2012a, Christensen2012b}, 2: \citet{Steidel2014}, 3: \citet{Erb2016}, 4: \citet{Fosbury2003}, 5: \citet{Villar-Martin2004}, 6: \citet{Yuan2009}, 7: \citet{Bayliss2014}, 8: \citet{Pettini2010}, 9: \citet{James2014}, 10: \citet{Stark2014a}, and 11: \citet{Steidel2016}.\\
\hbox to6pt{\footnotemark[a]\hss} Upper/Lower limit at the $2\sigma$ level.\\
\hbox to6pt{\footnotemark[b]\hss} Taken form \citet{Erb2016}.\\
\hbox to6pt{\footnotemark[c]\hss} Stellar mass estimated with the instantaneous burst model and the {\Hb} luminosity measurement \citep{Villar-Martin2004}.\\
\hbox to6pt{\footnotemark[d]\hss} We estimate that the continuum level of {\Lynxno} is 0.5 $\mu$Jy at the rest-frame wavelength of the {\Lya} emission line \citep[][Figure 5]{Fosbury2003}. The {\Lya} flux is given in Table 2 of \citet{Fosbury2003}. We calculate the EW$_0$({\Lya}) value for {\Lynxno}, and show the value in this table. 
Even if we make a conservative estimation of EW$_0$({\Lya}) with the continuum level of 3.0 $\mu$Jy, we obtain the rest-frame EW$_0$({\Lya}) value of 80 \AA.\\
\hbox to6pt{\footnotemark[e]\hss} Median values and 1 $\sigma$ scatters of the 30 KBSS galaxies.\\
\hbox to6pt{\footnotemark[f]\hss} The EW$_0$({\Lya}) value of the S16-stack is estimated with the continuum value of $\sim 0.3\mu$ Jy indicated from \citep[][Figure 2]{Steidel2016}
and the Ly$\alpha$ flux presented in Table 5 of \citet{Steidel2016}. The uncertainty of the continuum value does not significantly change the results.\\
\hbox to6pt{\footnotemark[$\dagger$]\hss} N/O values and upper limits obtained with Equations (\ref{eq:nppopp}) and (\ref{eq:no2}). The N/O values of the other objects are estimated with Equations (\ref{eq:npop}) and (\ref{eq:no1}).
}\label{tab:prop}
\end{table}
\end{landscape}

%% file: tab3.tex
\begin{landscape}
\begin{table}[!hbtp]
\caption{Forbidden and Semi-Forbidden Lines} %
\begin{center}
{\scriptsize
\begin{tabular}{p{2pt}lcccccccccc}
\hline\noalign{\vskip3pt} 
ID &
Object &
\OIIIsemi &
\OIIIsemi &
\OIIIsemi &
\NIIIsemi &
\OII &
\OIII &
\OIII &
\OIII &
\OIII &
\NII \\
 &
 &
1661 &
1666 &
1661+1666 &
1750 &
3727 &
4363 &
4959 &
5007 &
4959+5007 &
6584 \\
(1) &
(2) &
(3) &
(4) &
(5) &
(6) &
(7) &
(8) &
(9) &
(10) &
(11) &
(12) \\
\hline\noalign{\vskip3pt} 
1&
\COSMOSno  & 
$0.049\pm0.012$ &
$0.075\pm0.015$ &
$0.124\pm0.019$ &
$<0.042$ &
$<2.90$ &
\ldots &
$1.89\pm0.42$ &
$6.46\pm0.29$ &
$8.35\pm0.53$ & 
$<0.100$\\
2&
\SMACSno  & 
$0.014\pm0.003$ &
$0.023\pm0.004$ &
$0.037\pm0.007$ &
\ldots &
$2.10\pm0.02$ &
\ldots &
$1.339\pm0.004$\footnotemark[a]&
$4.714\pm0.007$\footnotemark[a]&
$6.053\pm0.011$\footnotemark[a]&
$0.116\pm0.005$\footnotemark[a]\\
3&
\BXIno  & 
\ldots &
\ldots &
$0.12\pm0.01$ &
\ldots &
$1.01\pm0.04$ &
\ldots &
\ldots &
$7.9\pm1.0$ &
$10.37\pm0.52$ &
$0.12\pm0.03$ \\
4&
\BXIIno & 
\ldots &
\ldots &
$0.14\pm0.02$ &
\ldots &
$0.90\pm0.03$ &
\ldots &
\ldots &
$6.4\pm0.3$ &
$8.7\pm0.4$ &
$0.15\pm0.03$ \\
5&
\BXIIIno  & 
\ldots &
\ldots &
$0.21\pm0.04$ &
\ldots &
$0.87\pm0.04$ &
\ldots &
\ldots &
$6.4\pm0.3$ &
$9.6\pm0.5$ &
$<0.07$ \\
6&
\Lynxno  & 
\ldots &
\ldots &
$0.56\pm0.04$ &
$0.18\pm0.02$ &
$<0.25$ &
\ldots &
$2.58\pm0.3$ &
$7.50\pm0.3$ &
$10.08\pm0.6$ &
\ldots \\
7&
\AbellYKno  & 
\ldots &
\ldots &
\ldots &
\ldots &
$1.11\pm0.3$ &
$0.27\pm0.10$ &
$1.98\pm0.3$ &
$6.45\pm0.3$ &
$8.43\pm0.6$ &
$<0.050$\footnotemark[b]\\
8&
\SGASno  & 
$0.018\pm0.007$ &
$0.051\pm0.007$ &
$0.069\pm0.014$ &
$0.014\pm0.005$ &
$0.79\pm0.02$\footnotemark[c] &
$<0.014$ &
$2.35\pm0.02$ &
$8.08\pm0.02$ &
$10.43\pm0.04$ &
\ldots \\
9&
\CSWAno  & 
\ldots &
$0.10\pm0.04$ &
$0.14\pm0.06$ &
\ldots &
$0.99\pm0.05$ &
\ldots &
$1.67\pm0.10$ &
$6.64\pm0.39$ &
$6.64\pm0.39$ &
$0.06\pm0.02$ \\
10&
\MACSno & 
$0.2\pm0.1$ &
$0.3\pm0.1$ &
$0.5\pm0.2$ &
$<0.1$ &
$<0.63$ &
\ldots &
$1.37\pm0.05$ &
$3.95\pm0.05$ &
$5.31\pm0.10$ &
$<0.065$ \\
11&
\AbellSno  & 
$<0.03$ &
$0.083\pm0.029$ &
$0.12\pm0.04$ &
$<0.03$ &
$<1.2$ &
\ldots &
$2.03\pm0.03$ &
$6.12\pm0.10$\footnotemark[d] &
$8.15\pm0.13$\footnotemark[d] &
\ldots \\
12 &
\KBSS  & 
$0.009\pm0.003$ &
$0.025\pm0.003$ &
$0.034\pm0.004$ &
\ldots &
$2.15\pm0.04$ &
$<0.06$ &
$1.46\pm0.02$ &
$4.37\pm0.02$ &
$5.83\pm0.02$ &
$0.35\pm0.02$ \\
\hline\noalign{\vskip3pt} 
\end{tabular}}
\end{center}
{\footnotesize 
(1): ID. (2): Name of the object. (3)--(12): Forbidden and semi-forbidden line fluxes normalized by the \Hb\ flux. The upper limits correspond to the 2$\sigma$ levels. The fluxes are \textit{not} corrected for dust extinction. \\
\hbox to6pt{\footnotemark[a]\hss} Under the influence of the telluric absorption lines. We do not correct the flux loss by the telluric absorption.\\
\hbox to6pt{\footnotemark[b]\hss} Including the potential uncertainty from the low instrumental response at the wavelength.\\
\hbox to6pt{\footnotemark[c]\hss} Under the influences of the poor atmospheric throughput and sky lines at the wavelength.\\
\hbox to6pt{\footnotemark[d]\hss} \OIII\,$\lambda$5007 flux estimated from the \OIII$\lambda$4959 flux measurement with the assumption of \OIII\,$\lambda$5007/$\lambda$4959 = 3.01 that is self-consistently calculated with our \textsc{Cloudy} model. \OIII\,$\lambda$5007 of this object is not detected, due to the strong telluric absorption.
}\label{tab:forbidden_lines}
\end{table}
\end{landscape}

%% file: tab4.tex
\begin{table*}[!hbtp]
\caption{Balmer Line Fluxes and the Color Excess}
\begin{center}
{\scriptsize
\begin{tabular}{p{3pt}lccccccc}
\hline\noalign{\vskip3pt} 
ID &
Object &
\Hd &
\Hc &
\Hb &
\Ha &
$E(B-V)_{\rm neb}$ &
Notes \\
 &
 &
4101 &
4340 &
4861 &
6563 &
(mag) &
 \\
(1) &
(2) &
(3) &
(4) &
(5) &
(6) &
(7) & 
(8) \\
\hline\noalign{\vskip3pt} 
1 &
\COSMOSno  & 
\ldots & 
\ldots & 
$1.0\pm0.3$ \footnotemark[\dag] & 
$2.84\pm0.05$ \footnotemark[\dag] &
$0.01^{+0.43}_{-0.01}$&
\ldots \\
2&
\SMACSno  & 
$0.257\pm0.005$\footnotemark[\dag] &
$0.412\pm0.003$\footnotemark[\dag] &
$1.000\pm0.002$\footnotemark[a]&
$3.604\pm0.001$\footnotemark[a]&
$0.00\pm0.00$& 
\ldots \\
3&
\BXIno  & 
\ldots &
\ldots &
$1.0$\footnotemark[\dag] &
$3.46\pm0.25$\footnotemark[\dag]  &
$0.17^{+0.06}_{-0.07}$&
\ldots \\
4&
\BXIIno  & 
\ldots &
\ldots &
$1.0$\footnotemark[\dag] &
$2.81\pm0.2$\footnotemark[\dag]  &
$0.00^{+0.06}_{-0.00}$&
\ldots \\
5&
\BXIIIno  & 
\ldots &
\ldots &
$1.0$\footnotemark[\dag] &
$2.77\pm0.2$\footnotemark[\dag]  &
$0.00^{+0.03}_{-0.00}$&
\ldots \\
6&
\Lynxno & 
\ldots &
\ldots &
$1.0$ &
\ldots  &
\ldots  & 
A\\
7&
\AbellYKno & 
\ldots &
\ldots &
$1.0\pm0.1$\footnotemark[\dag] &
$5.03\pm0.4$\footnotemark[b]\footnotemark[\dag]  &
$0.51^{+0.12}_{-0.10}$&
B\\
8&
\SGASno & 
\ldots &
$0.52\pm0.02$\footnotemark[c]\footnotemark[\dag] &
$1.00\pm0.01$\footnotemark[\dag] &
\ldots  &
$0.00\pm0.00$&
B\\
9&
\CSWAno  & 
\ldots &
$0.57\pm0.05$ &
$1.00\pm0.07$\footnotemark[\dag] &
$3.32\pm0.20$\footnotemark[\dag]  &
$0.14^{+0.08}_{-0.09}$&
\ldots \\
10&
\MACSno & 
\ldots &
\ldots &
$1.0\pm0.1$\footnotemark[\dag] &
$2.58\pm0.16$\footnotemark[\dag]  &
$0.00^{+0.04}_{-0.00}$&
\ldots \\
11&%
\AbellSno  & 
\ldots &
\ldots &
$1.00\pm0.03$\footnotemark[\dag] &
$2.97\pm0.03$\footnotemark[\dag] &
$0.04\pm0.03$&
\ldots \\
12&%
\KBSS  & 
\ldots &
\ldots &
$1.00\pm0.02$\footnotemark[\dag] &
$3.61\pm0.02$\footnotemark[\dag] &
$0.21\pm0.02$&
\ldots  \\
\hline\noalign{\vskip3pt} 
\end{tabular}}
\end{center}
{\footnotesize 
(1): ID. (2): Name of the object. (3)--(6): Fluxes of the Balmer lines normalized 
by \Hb\ emission-line fluxes. 
These Balmer line fluxes are \textit{not} corrected for dust extinction. 
(7): Color excess derived from the Balmer decrements. 
(8): ``A'' denotes a galaxy with no Balmer decrement measurement and $E(B-V)_{\rm neb}$.
``B'' indicates a galaxy with potential uncertainties that are originated from the strong sky lines or the low throughputs at the wavelengths of the Balmer lines. \\
\hbox to6pt{\footnotemark[a]\hss} Under the influence of telluric absorption lines.\\
\hbox to6pt{\footnotemark[b]\hss} Under the influence of the low instrumental response at the wavelength of \Ha.\\
\hbox to6pt{\footnotemark[c]\hss} Under the influences of the poor atmospheric throughput and the moderately strong sky lines.\\
\hbox to6pt{\footnotemark[$\dagger$]\hss} Balmer lines used for the $E(B-V)_{\rm neb}$ estimate.
}\label{tab:balmer_lines}
\end{table*}

%% file: tab5.tex
\begin{table}[!hbtp]
\caption{Balmer Decrements from the Model}
\begin{center}
{\scriptsize
\begin{tabular}{cccc}
\hline\noalign{\vskip3pt} 
$T_e$ &
\Ha/\Hb &
\Hc/\Hb &
\Hd/\Hb \\
(K) &
 &
 &
 \\
(1) &
(2) &
(3) &
(4) \\
\hline\noalign{\vskip3pt} 
5000& 
3.04&
0.458&
0.251\\
10000& 
2.86&
0.468&
0.259\\
20000& 
2.75&
0.475&
0.264 \\
\hline\noalign{\vskip3pt} 
\end{tabular}}
\end{center}
{\footnotesize 
Under the assumption of the Case B recombination with $n_e=100$ cm$^{-3}$ \citep{Osterbrock1989}.
(1): Electron temperature. (2)--(4): Balmer line ratios.
}\label{tbl:caseB}
\end{table}

%% file: tab6.tex
\begin{table}[!hbtp]
\caption{ISM Properties of Galaxies}
\begin{center}
{\scriptsize\
\begin{tabular}{lccc}
\hline\noalign{\vskip3pt} 
Objects &
{\q} excess &
N/O excess &
Class \\
(1) &
(2) &
(3) &
(4) \\
\hline\noalign{\vskip3pt} 
\multicolumn{4}{c}{\bf z$\sim$2 galaxies}\\ \hline
S16-stack &
{\it no} &
{\it yes} &
1) \\ 
{\COSMOS}\footnotemark[a], {\BXIII}, {\AbellS} &
{\it yes} &
{\it no} &
2) \\
{\BXI}, {\BXII} &
{\it yes} &
{\it yes} &
3) \\ \hline
\multicolumn{4}{c}{\bf local galaxies}\\ \hline
Reference galaxy &
{\it no} &
{\it no} &
0) \\
3 LMHSs
and 1 GP &
{\it no} &
{\it yes} &
1) \\
3 GPs&
{\it yes} &
{\it no} &
2) \\
1 GP&
{\it yes} &
{\it yes} &
3) \\
\hline\noalign{\vskip3pt} 
\end{tabular}}
\end{center}
{\footnotesize 
(1): IDs or populations of Galaxies. 
(2): With and without a {\q} excess for "yes" and "no", respectively.
(3): With and without an N/O excess for "yes" and "no", respectively. 
(4): Classifications; 
0) for galaxies with {\it neither} N/O nor {\q} excess,
1) for galaxies with an N/O excess and {\it no} {\q} excess,
2) for galaxies with a {\q} excess and {\it no} N/O excess, and
3) for galaxies with both N/O and {\q} excesses.\\
\hbox to6pt{\footnotemark[a]\hss} A possible {\q} excess indicated by the lower limit on {\q}.
}\label{tab:class}
\end{table}